%% file: main.tex
\documentclass[10pt,twocolumn,showpacs,superscriptaddress,aps,prl,floatfix]{revtex4-2}
\usepackage{graphicx}
\usepackage{amsmath,amssymb}
\usepackage[utf8]{inputenc}
\usepackage[T1]{fontenc}
\usepackage{lmodern}
\usepackage{natbib}
\usepackage{hyperref}
\usepackage{color}
\hypersetup{colorlinks=true,citecolor={blue},linkcolor={blue},urlcolor={blue}}
\usepackage{units}
\usepackage[english]{babel}
\usepackage{dcolumn} 
\usepackage{bm}
\usepackage{xfrac}
\usepackage[normalem]{ulem}
\usepackage{nicefrac}
\usepackage{bm} 
\usepackage{bbm}
\usepackage{soul}
\setstcolor{red}
\usepackage{xcolor,cancel}
\usepackage[capitalize]{cleveref}
\usepackage{tikz}
\usepackage{physics}
\usepackage[caption=false,lofdepth,lotdepth]{subfig}

\usepackage{soul}

\def\(({\left(}
\def\)){\right)}
\newcommand{\be}{\begin{equation}}
\newcommand{\ee}{\end{equation}}
\newcommand{\bea}{\begin{eqnarray}}
\newcommand{\eea}{\end{eqnarray}}

\newcommand{\s}{\sigma}

\newcommand{\Acal}{\mathcal{A}}
\newcommand{\Bcal}{\mathcal{B}}
\newcommand{\Ccal}{\mathcal{C}}
\newcommand\sinc {\mathrm{sinc}}

\begin{document}


\title{Encoding arbitrary Ising Hamiltonians on Spatial Photonic Ising Machines}

\author{Jason Sakellariou}\email{isakellariou@q.ubitech.eu}
\affiliation{QUBITECH, Thessalias 8, GR 15231 Chalandri, Athens, Greece}
\affiliation {UBITECH Ltd., 95B Archiepiskopou Makariou, CY 3020 Limassol Cyprus}
\author{Alexis Askitopoulos}\email{aaskitopoulos@q.ubitech.eu}
\affiliation{QUBITECH, Thessalias 8, GR 15231 Chalandri, Athens, Greece}
\affiliation {UBITECH Ltd., 95B Archiepiskopou Makariou, CY 3020 Limassol Cyprus}
\author{Georgios Pastras}
\affiliation{QUBITECH, Thessalias 8, GR 15231 Chalandri, Athens, Greece}
\author{Symeon I. Tsintzos}\email{stsintzos@q.ubitech.eu}
\affiliation{QUBITECH, Thessalias 8, GR 15231 Chalandri, Athens, Greece}
\affiliation {UBITECH Ltd., 95B Archiepiskopou Makariou, CY 3020 Limassol Cyprus}

\date{\today}

\begin{abstract}
    Photonic Ising Machines constitute an emergent new paradigm of computation,
    geared towards tackling combinatorial optimization problems that can be
    reduced to the problem of finding the ground state of an Ising model.
    Spatial Photonic Ising Machines have proven to be advantageous for
    simulating fully connected large-scale spin systems. However, fine control
    of a general interaction matrix $J$ has so far only been accomplished
    through eigenvalue decomposition methods that either limit the scalability
    or increase the execution time of the optimization process. We introduce
    and experimentally validate a SPIM instance that enables direct control
    over the full interaction matrix, enabling the encoding of Ising
    Hamiltonians with arbitrary couplings and connectivity. We
    demonstrate the conformity of the experimentally measured Ising
    energy with the theoretically expected values and then proceed to solve
    both the unweighted and weighted graph partitioning problems, showcasing a
    systematic convergence to an optimal solution via simulated annealing. Our
    approach greatly expands the applicability of SPIMs for real-world
    applications without sacrificing any of the inherent advantages of the
    system, and paves the way to encoding the full range of NP problems that
    are known to be equivalent to Ising models, on SPIM devices.
\end{abstract}
\maketitle

In recent years, a growing effort has been devoted towards the implementation
of special purpose physical machines that can simulate the Ising Hamiltonian.
These machines have garnered considerable attention for their potential to
efficiently tackle optimization problems across diverse domains, as many
non-deterministic polynomial-time hard (NP-hard) problems can be mapped to the
Ising Hamiltonian~\cite{mezard1987spin, mezard2009information,
lucas_ising_2014}. A number of physical implementations of Ising machines,
employing either quantum or classical schemes, have been developed. Quantum
Ising annealers have been realized by trapped atoms~\cite{scholl_quantum_2021}
and ions~\cite{kim_quantum_2010}, single photons~\cite{ma_quantum_2011} and
superconducting circuits~\cite{johnson_quantum_2011}. In addition, their
classical counterparts have been realized in various physical platforms
including polariton
condensates~\cite{berloff_realizing_2017,ohadi_spin_2017,kalinin_polaritonic_2020,alyatkin_optical_2020},
stochastic magnetic junctions~\cite{borders_integer_2019},
memristors~\cite{cai_power-efficient_2020}, coupled electrical
oscillators~\cite{shukla_synchronized_2014}, complementary metal oxide
semiconductor technologies (CMOS)~\cite{merolla_million_2014}, networks of
coupled optical pulses in a ring fiber~\cite{mcmahon_fully_2016,
inagaki_coherent_2016} and lately spatial photonic Ising machines
(SPIMs)~\cite{pierangeli_large-scale_2019}.

Photonic Ising machines are of particular interest as they exploit the
advantages of mature optoelectronic techologies and have been shown to address
large-scale combinatorial optimization
problems~\cite{honjo_100000-spin_2021,pierangeli_large-scale_2019}. SPIMs, a
newcomer to the field, utilize holographic optical phase modulation, leveraging
1) enhanced scalability, 2) all-to-all connectivity, 3) room temperature
operation, 4) inherent parallelism 5) low cost and 6) low power consumption.
Compared to other photonic Ising machines, SPIMs utilize a straightforward
optical configuration to encode the physical parameters of the Ising
Hamiltonian as distinct holographic phases on a discretized optical wavefront
with the use of spatial light modulators (SLM)~\cite{sun_quadrature_2022}.
Since their introduction in 2019~\cite{pierangeli_large-scale_2019}, SPIMs have
showcased their applicability to adiabatic evolution
methods~\cite{pierangeli_adiabatic_2020}, while the inherent system noise was
shown to be a valuable system resource guiding the convergence towards
low-energy states~\cite{pierangeli_noise-enhanced_2020}. Additionally, more
advanced schemes utilizing optical non-linearities have also been implemented
to include four-body interaction
terms~\cite{kumar_large-scale_2020,kumar_observation_2023}. Furthermore, they
have been used to study the thermodynamics of Ising
systems~\cite{fang_experimental_2021,leonetti_optical_2021}, and for tackling
various NP-hard  problems such as the Number Partitioning
Problem~\cite{huang_antiferromagnetic_2021,prabhakar_optimization_2022},
Max-Cut~\cite{ye_photonic_2023} and Knapsack
problems~\cite{yamashita_low-rank_2023}. While several computationally
interesting problems can already be handled by
SPIMs~\cite{wang_efficient_2024}, the standard SPIM configuration is restricted
to Mattis type interactions~\cite{mattis_solvable_1976}, a fact that poses a
significant limitation. Various approaches, such as vector matrix
multiplication~\cite{ouyang_-demand_2024}, time
division~\cite{wang_general_2024} and wavelength multiplexing
schemes~\cite{luo_wavelength-division_2023}, have been suggested to overcome
this constraint, by decomposing the interaction matrix $J$ in a series of
Mattis models that can then be independently treated with a SPIM, limiting
however, the scalability of the system or significantly increasing the
execution time of the optimization process.

In this letter we formulate an alternative spin-interaction encoding for SPIMs
that allows for the manipulation of arbitrary coupling matrices $J$. Moreover,
this method offers scaling advantages in the case of sparse models. Each term
$J_{ij}\sigma_{i}\sigma_{j}$ of the Ising Hamiltonian is directly encoded as an
element of the phase matrix imprinted on the SLM. We validate our method
through a direct comparison of the theoretically expected to the experimentally
measured energy levels of a random spin glass Hamiltonian. Then, we apply this
scheme to the Graph Partitioning Problem (GPP), showcasing persistently
high-quality solutions for problems of arbitrary sparsity. Finally, we expand
our approach to the weighted version of the problem, where the elements of the
coupling matrix $J_{ij}$ take random positive values.

\section*{Spin-Product-Encoding of arbitrary Ising Hamiltonians}

The existing approach to the treatment of Ising models by
SPIMs~\cite{pierangeli_large-scale_2019} is based on the correspondence of each
SLM pixel, which is a square with width $L$, to a spin of the Ising model. Let $\zeta_i$ be
the amplitude of the incoming electric field to the $i$-th pixel. The SPIM adds
a phase $\varphi_i$ to the electric field. In the simplest approach this phase
is equal to the value of the corresponding spin, but in principle it may also
contain a constant angle $\theta_i$ different for each pixel, i.e. $\varphi_i =
\sigma_i e^{i \theta_i}$. For the moment let us assume that we do not use this
extra freedom, i.e. $\theta_i = 0$.

For a given spin configuration $\s = \{ \s_1, \dots, \s_N \}$, the
corresponding SLM pixels are set to the appropriate phases. The electric field
is then Fourier transformed by a Fourier lens and projected onto the camera.
The captured image is the intensity
\begin{equation}
    \widetilde{I} \left( \vec{k} \right) = \sinc^2 \frac{k_x L}{2} \sinc^2 \frac{k_y L}{2} \sum_{i, j} \zeta_i \zeta_j^* \varphi_i \varphi_j^* e^{i \vec{k} \cdot \left( \vec{r}_i - \vec{r}_j \right)} .
\label{eq:electric_intensity}
\end{equation}

The Ising energy is computed as
${H(\s)=-\widetilde{I}(\vec{0})}$, which yields
\begin{equation}
    H(\s) = -  \Re \sum_{i,j} \zeta_i \zeta_j^* \varphi_i \varphi_j^* .
    \label{eq:Hamiltonian_phases}
\end{equation}
The relation between the phases $\varphi_i$ and the values of the spins implies
that the above Hamiltonian is simply the Ising model
\begin{equation}
    H(\s) = - \sum_{i,j} J_{ij} \sigma_i \sigma_j ,
\end{equation}
with couplings
\begin{equation}
    J_{ij} =  \Re \zeta_i \zeta_j^* .
\end{equation}
Further assuming that the phase of the incoming electric field is uniform, i.e.
the amplitudes $\zeta_i$ are real, this corresponds to the Mattis model
${H\propto-\sum_{i,j}\zeta_i\zeta_j\s_i\s_j}$. This approach  results in a
restricted class of models since the number of controllable parameters (the
amplitudes $\zeta_i$) is linear to the number of spins $N$, whereas the general
Ising model has a number of independent couplings that scales as $O(N^2)$.

The main idea of this work is the following: Being able to encode individual
binary spins using an SLM implies that one can also encode products of spins,
as they also take values $\pm 1$. It follows that SLM pixels can be assigned to
products of spins instead of single spins.
In what follows this method is called \textit{Spin-Product-Encoding} (SPE).
This idea can be particularly handful
in the case of Ising models defined on sparse graphs, due to the fact that only
non-zero couplings are allocated to SLM pixels.

Let the Ising model be defined on an interaction graph $G=(V,E)$, with $N=|V|$
spins identified by the index $i$, and non-vanishing couplings only between the
pairs of spins ${(i,j)\in E}$. The Ising Hamiltonian reads
\begin{equation}
    H(\s) = - \sum_{(i , j) \in E} J_{ij} \sigma_i \sigma_j ,
    \label{eq:H_Ising}
\end{equation}

We match each pixel to a pair of spins $(i , j) \in E$. These pixels add a
phase delay to the electric field equal to ${\varphi_{ij} = \sigma_i \sigma_j}$.
We furthermore employ an ancillary spin $\s_0$, assigned to one SLM pixel.
Assuming real incoming electric field amplitudes $\zeta_{ij}$ and $\zeta$ for the set of
pixels assigned to spin pairs and the ancillary spin respectively, the Hamiltonian in
\cref{eq:Hamiltonian_phases} reads
\begin{multline}
    \widetilde{H}(\s_0, \s) = -  \zeta \s_0 \sum_{(i , j) \in E} \zeta_{ij} \s_i \s_j \\
    -  \left( \sum_{(i , j) \in E} \zeta_{ij} \s_i \s_j \right)^2 - \zeta^2 .
    \label{eq:Hamiltonian_tilde}
\end{multline}

The first term of \cref{eq:Hamiltonian_tilde} is proportional to the Hamiltonian of a
generic Ising model with couplings
\begin{equation}
    J_{ij} = \zeta \zeta_{ij} .
    \label{eq:edge_couplings}
\end{equation}
The last term is just a non-dynamical constant. The second term, however,
involves four-spin couplings. In order to cancel the four-spin contributions
we obtain two measurements of $\widetilde{H}$, for the two
values of $\s_0$, while keeping the rest of the spins fixed. Then we have
\begin{equation}
    H(\s) =\frac{\widetilde{H}(+1, \s) - \widetilde{H}(-1, \s)}{2}  = - \zeta \sum_{(i, j) } \zeta_{ij} \sigma_i \sigma_j ,
    \label{eq:Hamiltonian_pm}
\end{equation}
which is the generic Ising model with couplings given by \cref{eq:edge_couplings}.

The treatment of an arbitrary
Ising model with this method requires the modulation of the incoming laser
beam, since the couplings $J_{ij}$ are determined by the amplitudes
$\zeta_{ij}$. We can trade amplitude modulation for phase modulation, which can
be achieved by the SLM pixels, as shown in the supplementary material~\cite{supp}.

\section*{Experimental configuration}

\begin{figure}
    \centering
    \def\svgwidth{0.48\textwidth}
    \input{schematic.tex}

    \setcounter{subfigure}{1}
    \subfloat[]{
        \resizebox{0.49\textwidth}{!}{
            {\Huge
                \ensuremath{
                    \raisebox{-0.465\height}{\includegraphics[width=5cm]{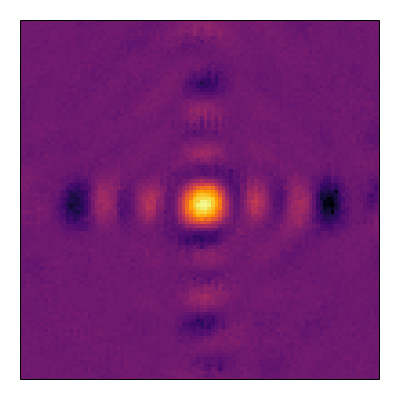}} =
                    \frac{1}{2} \left( \raisebox{-0.465\height}{\includegraphics[width=5cm]{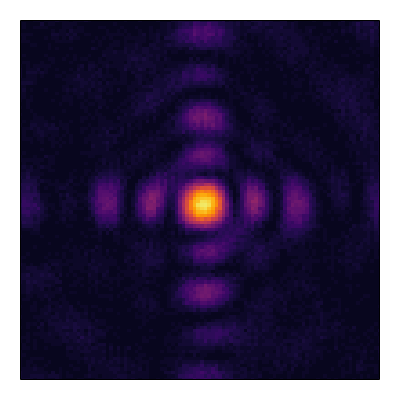 }} -
                    \raisebox{-0.465\height}{\includegraphics[width=5cm]{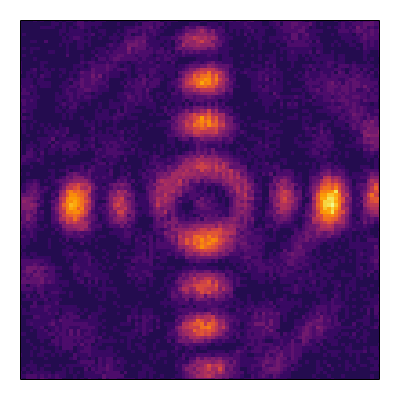 }} \right)
                }
            }
        }
    }

    \vspace{8pt}
    \resizebox{0.49\textwidth}{!}{\input{energy_scatter.tex}}
    \caption{
        (a) Schematic of the experimental setup. (b) Images captured by the camera
        corresponding to the energy computation according to \cref{eq:Hamiltonian_pm} for
        a given spin configuration. (c,d) The experimental energies
        versus the theoretical ones for a sparse ferromagnetic system (c), and a
        sparse spin glass system (d).
    }
    \label{fig1}
\end{figure}

The optical configuration of our SPIM is depicted in \cref{fig1}(a). Light from
a stabilised continuous wave He\nobreakdash-Ne laser is expanded 10 times and
impinges on a reflective spatial light modulator (SLM), Holoeye PLUTO-2.1-NIR.
In addition to the dynamic spin phase encoding, a static holographic grating is
applied to separate the 1st diffracted order from the unmodulated reflected
light. The beam is then focused on a high QE Peltier cooled CMOS camera, Atik
Camera ACICS 7.1. The region of interest of the camera is selected around the
1st order of the holographic grating. To compensate camera intensity variations
due to laser fluctuations and SLM flickering, we record the modulated laser
light after impinging the SLM, by placing a power meter (PM). Then, the recorded
image is normalized with the corresponding power value captured by the PM.

Since the SPIM Ising energy is directly related to the recorded light
intensity, a calibration process takes place at the beginning of each
experimental run by comparing the experimental energy values to the
corresponding theoretical ones. First we sample random spin configurations,
uniformly distributed in the whole configuration space. Then we perform a
linear fit between the experimental and theoretical energies, and we obtain the
normalization factor and offset for the experimental ones.

\Cref{fig1}(c,d) presents the results of the above calibration process for a
sparse ferromagnetic and a sparse spin glass system, showing a perfect match
between the corresponding experimental and theoretical energies. The
distribution of $J_{ij}$ for the case of the sparse ferromagnet is $P(J) = p
\delta(J-1) + (1-p) \delta(J)$, while for the spin glass is $P(J) = p
\mathbbm{1}_{[-1,1]}(J) + (1-p) \delta(J)$, where $\mathbbm{1}_{[-1, 1]}$ is
the uniform distribution in the interval $[-1,1]$ and $p$ is the \textit{edge probability}
of the graph. We used $p=0.05$ for both cases in \cref{fig1}(c,d).

\section*{Graph Partitioning}

We apply our approach to the GPP, which is a
member of the NP-hard complexity class. This problem is suitable in order to
demonstrate the ability of our method to tackle problems with arbitrary
sparsity, overcoming the limitations of existing solutions.

The GPP considers an undirected graph $G=(V,E)$ with an even number $N=|V|$ of
vertices. The problem asks for a partition of the set $V$ into two subsets
$\Acal$ and $\Bcal = V \setminus \Acal$ of equal size,
such that the number of edges connecting the two subsets is minimized.
The \textit{cut-set} is defined as $\Ccal = \{ (i,j) \in E \mid i \in \Acal, j \in \Bcal \}$,
and the \textit{cost} for the GPP as the size of the cut-set $C = |\Ccal|$.

Following~\cite{lucas_ising_2014} the GPP can be mapped to the Ising
model using the following Hamiltonian
\begin{subequations}
    \be
        H = a H_a + b H_b, \\ \text{ with}
        \label{eq:H_GPP_ΑΒ}
    \ee
    \be
        H_a = \left(\sum_{i=1}^{N}\sigma_{i} \right)^2, \\
        \label{eq:H_GPP_A}
    \ee
    \be
        H_b = \sum_{ (i, j) \in E}\frac{1-\sigma_i \sigma_j}{2}.
        \label{eq:H_GPP_B}
    \ee
    \label{eq:H_GPP}
\end{subequations}
The term $H_a$ penalises partitions into subsets of unequal size, while the
term $H_b$ corresponds precisely to the cost $C$. In
this work we use $a=b=1$. The mapping between the GPP and the above Hamiltonian
works as follows. Once the ground state of the Hamiltonian is found, the graph
vertices are assigned to the two subsets according to the sign of the
corresponding spins, i.e. ${\Acal=\{i \mid \sigma_i=1\}}$ and ${\Bcal=\{i \mid \sigma_i=-1\}}$.

In order to solve the GPP we implement the Hamiltonian of \cref{eq:H_GPP} on
the SPIM using a hybrid encoding scheme. We implement $H_a$ using the existing
approach found in~\cite{pierangeli_large-scale_2019} and $H_b$ using the new
encoding scheme proposed in this work. The term $H_a$ has homogeneous couplings
and is fully connected, which is the ideal case for the existing method, as the
number of pixels required scales linearly with the number of spins. On the
other hand, the term $H_b$ has sparse couplings, which is the ideal case for
the new encoding scheme, as the number of pixels required scales linearly with
the number of edges of the interaction graph $|E|$. The two terms are computed
sequentially in two steps, and the final energy is obtained by summing the two
contributions.

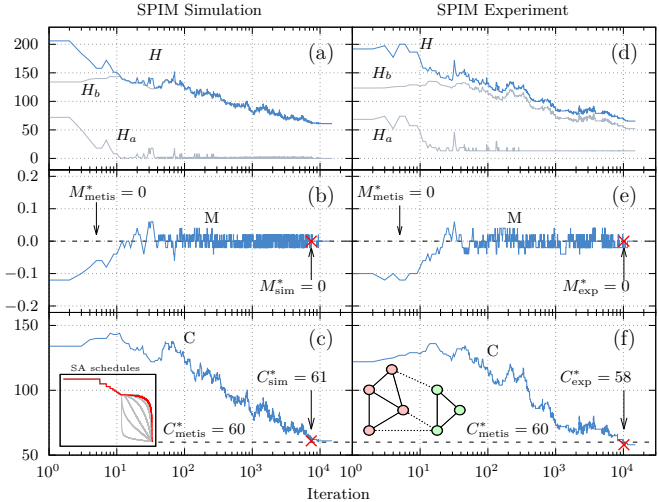
\begin{figure}
\centering
    \resizebox{0.48\textwidth}{!}{\input{graph_part_sim_and_spim.tex}}
    \caption{
        Simulated (a,b,c) and experimental (d,e,f) results for
        a graph partitioning Hamiltonian, showing the total ($H$) and
        individual ($H_a$ and $H_b$) energies (a,d), the magnetization (b,e),
        and the cost (c,f) as a function of the optimization iterations. Inset
        in (c) depicts the different annealing schedules that were tried (grey)
        and the one that was used (red). Inset in (f) shows a graphical
        representation of a toy example of the graph partitioning problem. The
        dashed lines in (b,c,e,f) represent the solution obtained by
        the METIS package.
    }
    \label{fig2}
\end{figure}

\Cref{fig2} illustrates the results obtained from both simulation
and experimental realization by applying the proposed optical encoding scheme
to the GPP. In this case, we considered a graph with $N=100$ vertices and a
edge density of $p=0.05$. In addition, we used the software package
METIS~\cite{karypislabmetis_2024} that implements a state-of-the-art algorithm
for GPP ~\cite{lasalle_parallel_2016} to compare with our results.

Both the simulated and experimental results are obtained by running a Simulated
Annealing (SA) optimization algorithm to minimize the Hamiltonian in
\cref{eq:H_GPP}. Various annealing schedules were implemented and tested, as
shown in the inset of \cref{fig2}(c). Among them, the
\textit{linear additive} (red line), was selected because we found empirically
that it provides the best results.

The simulation results were obtained by developing a model based on Fresnel
diffraction theory. The model incorporates the optical encoding of the spins
and interactions among them as additional phase masks, mimicking the role of
the SLM.

The top panel of \cref{fig2}(a,c) presents the evolution of the total energy,
$H$, and the two individual energy terms, $H_a$ and $H_b$, as defined in
\cref{eq:H_GPP}, as a function of the iteration number of  the SA algorithm. After a
number of iterations, the SA algorithm converges to a minimum for the total
energy. The cost $C$, obtained by the simulated and experimental solution of
the GPP is shown in \cref{fig2}(c,f). The red cross on the graphs
denotes the \textit{optimal cost} $C^*$ obtained from the minimization
procedure, while the dashed line shows the solution of the METIS package. In
particular, the simulation and experimental values are $C^*_{\text{sim}}=61$ and
$C^*_{\text{exp}}=58$, respectively, while $C^*_{\text{metis}}=60$.

The aforementioned simulation and experimental values were obtained by
selecting the spin configuration that minimizes the cost while having vanishing
magnetization, see \cref{fig2}(b,e). This means that we are only interested in
solutions that split the graph into two equal subsets, as required by the GPP.
Both values are comparable to the METIS solution, while the experimental value
slightly outperforms both simulation and METIS, highlighting the versatility of our
method to address arbitrary graphs using the SPIM optical architecture.

\begin{figure}
    \centering
        \resizebox{0.49\textwidth}{!}{\input{stats-gpp-wgpp.tex}}
        \caption{
            Multiple runs obtained experimentally for the Unweighted Graph
            Partitioning Problem (GPP) (a,b) and the Weighted Graph
            Partitioning Problem (WGPP) (c,d). Both instances have $N=100$ and
            $p=0.05$. Each column shows the total energy (a,c) and the
            cost (b,d) for 10 (a,b) and 5 (c,d) different
            initial spin configurations, showing that the SPIM systematically
            finds solutions with comparable cost to that obtained by METIS.
            The insets show all individual runs.
        }
    \centering
    \label{fig3}
\end{figure}
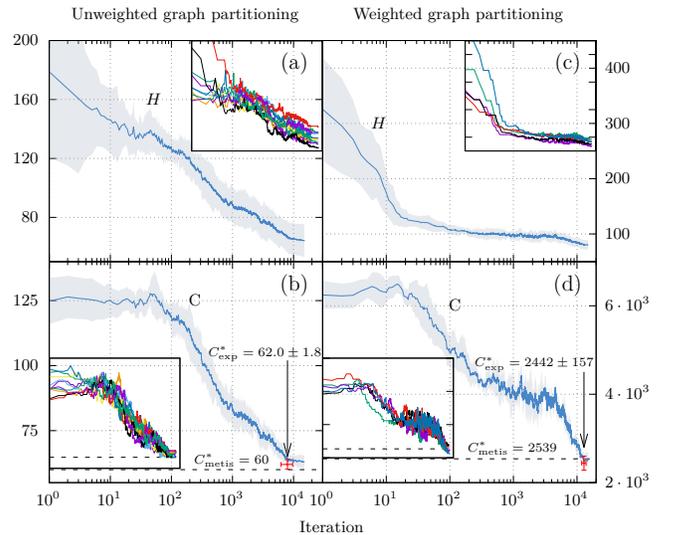

For the graph appearing in \cref{fig2}, we performed 10 distinct experimental
trials starting from different initial spin configurations,
to showcase the systematic convergence to a low-energy state.
The average energy
and cost obtained from this process are depicted in \cref{fig3}(a,b). The
average of the optimal cost is $C^*_{\text{exp}}=62\pm1.8$ and is depicted by the red cross.
For reference, the METIS solution is $C^*_{\text{metis}}=60$.

To demonstrate the ability of the method to handle general Ising models as
well, we proceed with the \textit{weighted} version of the GPP (WGPP). In WGPP,
the graph is defined as the triplet $G=(V,E,w)$, where
${w_{ij}\in\mathbb{N}^*}$ is the weight of the edge $(i,j)$. The objective in
this case is to partition the vertices into two subsets minimizing the cost
${C=\sum_{(i,j)\in\Ccal}w_{ij}}$. For the WGPP the $H_b$ term of
\cref{eq:H_GPP} takes the form
\be
    H_b^{(\text{weighted})} = \sum_{(i,j) \in E} w_{ij} \frac{1-\sigma_i \sigma_j}{2}.
\ee
Following the theoretical description provided in the supplementary material~\cite{supp},
we encode the $w_{ij}$
using two adjacent pixels with ${\theta_{ij}=\arccos(w_{ij}/w_{\text{max}})}$.
\Cref{fig3}(c,d) depicts the results obtained from a graph with $|V|=100$,
$p=0.05$. The $w_{ij}$ are uniformly random integers in the range
$[1,w_{\text{max}}]$ with $w_{\text{max}}=100$.

The results are consistent with the unweighted case. Our method successfully
finds solutions that are comparable to the ones obtained by the METIS package.

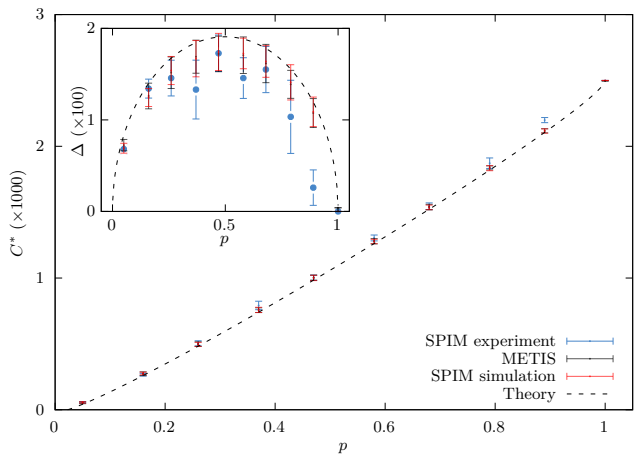
\begin{figure}
    \centering
        \resizebox{0.49\textwidth}{!}{\input{solution_vs_density.tex}}
        \caption{
            Optimal cost $C^*$ of the GPP for varying edge probability $p$ from
            the SPIM experiment, METIS, SPIM simulation and theory.
            The SPIM experiment points have been averaged over 20 random problem instances.
            The METIS and SPIM simulation points have been averaged over 50
            problem instances. Inset: the improvement due to optimization
            $\Delta(p)$.
        }
    \centering
    \label{fig4}
\end{figure}

Apart from the ability of the method to address arbitrary Ising models, it also
scales well for sparse graphs. The number of free parameters of the Ising model
scales as $O(N^2)$. However, for sparse graphs, a large number of these
parameters are zero, hence they do not contribute to the energy of the system.
By encoding only the non-zero couplings, our method requires a number of SLM
pixels that scales as $O(|E|)$.

To quantify the ability of the method to address sparse graphs, we
performed experiments to solve the GPP for various values of the edge
probability $p$. The results of those experiments are presented in \cref{fig4}
where the optimal cost is plotted as a function of $p$. In addition to the
experimental realizations, \cref{fig4} includes the results obtained from the
simulation model, the METIS package and the theoretical prediction for the
optimal cost obtained by the replica method~\cite{fu1986application,
mezard1987spin}, namely
\be
    C^*(p) = \frac{N^2}{4} p - 0.38 N^{3/2} \sqrt{p (1-p)}.
    \label{eq:optimal_cost}
\ee
Since the first term in \cref{eq:optimal_cost} does not depend on the
optimization~\cite{fu1986application}, we present in the inset of \cref{fig4}
the improvement due to optimization ${\Delta=pN^2/4-C^*}$. It is noteworthy
that the agreement between the simulation and METIS is perfect, and almost
matches the theoretical optimum. The agreement of the experimental results with
the rest is satisfactory, especially for sparse graphs, i.e. small values of
$p$.

\section*{Conclusion}

We have introduced and experimentally validated a novel encoding scheme for
SPIMs, effectively augmenting their functionality and applicability without
sacrificing scalability or speed. This allows for the direct encoding of the
full interaction matrix $J$ of any Ising Hamiltonian through a two-step iteration
process. The method requires $O(|E|)$ pixels,
hence it is particularly advantageous in the case of sparse graphs
in comparison to existing implementations~\cite{supp}. In the worst-case, $|E| \sim N^2$,
the method becomes equivalent to the existing implementations.

We applied this method on our SPIM instance to solve the unweighted
and weighted graph partitioning problems demonstrating comparable quality
solutions with a GPP specific algorithm for varying degree of sparsity.
Notably, our method can be easily expanded to Hamiltonians having
$p$\nobreakdash-spin interactions~\cite{supp}, which play an equally important
role in combinatorial optimization~\cite{bashar_designing_2023}. As the SLM
technology enters the MHz range~\cite{park_all-solid-state_2021}, SPIMs are
poised to become a powerful and versatile technology for real-world
applications.

\begin{acknowledgments}
We acknowledge useful discussions with C. Conti and N. Berloff. J.S., A.A., G.P.
and S.T. acknowledge support from HORIZON EIC-2022-PATHFINDERCHALLENGES-01
HEISINGBERG Project 101114978. J.S, A.A. and S.T. acknowledge support from
HoloCIM (CODEVELOP-ICTHEALTH/0322/0047).
\end{acknowledgments}

\end{document}


\title{Supplemental material: Encoding arbitrary Ising Hamiltonians on Spatial Photonic Ising Machines}

\author{Jason Sakellariou}
\affiliation{QUBITECH, Thessalias 8, GR 15231 Chalandri, Athens, Greece}
\affiliation {UBITECH Ltd., 95B Archiepiskopou Makariou, CY 3020 Limassol Cyprus}
\author{Alexis Askitopoulos}
\affiliation{QUBITECH, Thessalias 8, GR 15231 Chalandri, Athens, Greece}
\affiliation {UBITECH Ltd., 95B Archiepiskopou Makariou, CY 3020 Limassol Cyprus}
\author{Georgios Pastras}
\affiliation{QUBITECH, Thessalias 8, GR 15231 Chalandri, Athens, Greece}
\author{Symeon I. Tsintzos}
\affiliation{QUBITECH, Thessalias 8, GR 15231 Chalandri, Athens, Greece}
\affiliation {UBITECH Ltd., 95B Archiepiskopou Makariou, CY 3020 Limassol Cyprus}

\maketitle

\section{Trading amplitude modulation for phase modulation}

The treatment of an arbitrary
Ising model requires the modulation of the incoming laser
beam, since the couplings $J_{ij}$ are determined by the amplitudes
$\zeta_{ij}$. Here we show how we can trade amplitude modulation for phase modulation.
This is convenient since the phase modulation can be achieved by a single SLM.
This can be done by doubling the set of pixels
corresponding to the spin couplings. We distinguish the pixels corresponding to
the same spin coupling using the upper index $(\pm)$. It is essential to take
advantage of the extra spin-independent phase delay that can be introduced by
the SLM pixels. Namely, we introduce them so that obey $\theta_{ij}^{+} = -
\theta_{ij}^{-} \equiv \theta_{ij}$. In other words, $\varphi_{ij}^{\pm} =
\sigma_i \sigma_j e^{\pm i \theta_{ij}}$. The last part of the SLM is again
composed by one pixel that causes no phase delay to the electric field. No
modulation of the incoming electric field is applied, i.e. $\zeta_{ij}^{+} =
\zeta_{ij}^{-} = \zeta$. It is a matter of simple algebra to show that the
Hamiltonian in Eq.~(7) in the main text reads
\begin{equation}
    \widetilde{H}(\s_0, \s) = - \zeta^2 \s_0 \sum_{(i, j) \in E} \cos \theta_{ij} \sigma_i \sigma_j
    - 4 \zeta^2 \left( \sum_{(i, j) \in E} \cos \theta_{ij} \sigma_i \sigma_j \right)^2 - \zeta^2 .
    \label{eq:Hamiltonian_phases_nomod}
\end{equation}

Repeating the same procedure as described in the main text,
we obtain two measurements of $\widetilde{H}$,
then using Eq.~(8) in the main text we find
\begin{equation}
    H(\s) = -  \zeta^2 \sum_{(i, j) \in E} \cos \theta_{ij} \sigma_i \sigma_j ,
    \label{eq:Hamiltonian_phases_nomod2}
\end{equation}
which is the generic Ising model with couplings
\begin{equation}
    J_{ij} =  \zeta^2 \cos \theta_{ij} ,
\end{equation}
i.e. we managed to tune all couplings $J_{ij}$ without modulating the laser
beam, but solely with the use of the SLM, via the choice of appropriate values
for the angles $\theta_{ij}$.

\section{Extending the encoding scheme to the $p$-spin model}

Let us consider a model with $p$-spin interactions, i.e.
\begin{equation}
    H(\s) = - \sum_{1 \leq i_1 < \dots < i_p \leq N} J_{i_1 \dots i_p} \s_{i_1} \dots \s_{i_p} .
    \label{eq:Hamiltonian_pspin}
\end{equation}

This Hamiltonian can be encoded in the exact same way as the Hamiltonian in Eq.~(3) in the main text
that contains only 2-spin interactions. Each pixel is associated to a unique
product of $p$ spins, $\sigma_1 \sigma_2 \ldots \sigma_p$, that has a
non-vanishing coefficient in the Hamiltonian \eqref{eq:Hamiltonian_pspin}. Let
the incoming field amplitude to this pixel be $\zeta_{i_1 , i_2 , \ldots ,
i_p}$. We also introduce an ancillary spin $\sigma_0$ that is associated to a
single pixel of the SLM, where the incoming electric field amplitude is
$\zeta$. Finally, we configure the SLM pixels so that they add a phase delay to
the electric field equal to $\sigma_1 \sigma_2 \dots \sigma_p$.
It directly follows that the cost functional, i.e. the intensity measured at the center of the Fourier space reads
\begin{equation}
    \widetilde{H}(\s_0, \s) = -  \zeta \s_0 \sum_{1 \leq i_1 < \dots < i_p \leq N} \zeta_{i_1 \dots i_p} \s_{i_1} \dots \s_{i_p}
    -  \left( \sum_{1 \leq i_1 < \dots < i_p \leq N} \zeta_{i_1 \dots i_p} \s_{i_1} \dots \s_{i_p} \right)^2 - \zeta^2 ,
    \label{eq:Hamiltonian_tilde_pspin}
\end{equation}
Fixing the amplitudes so that
\begin{equation}
J_{i_1 , i_2 , \ldots , i_p} = \zeta \zeta_{i_1 , i_2 , \ldots , i_p}
\end{equation}
allows the encoding of the Hamiltonian \eqref{eq:Hamiltonian_pspin} via two
iterations, using Eq.~(8) in the main text.

The incoming amplitude modulation can be avoided by doubling the size of the
SLM, via the same mechanism as the one described in the previous section.
Let us note that a model with local magnetic fields of the form
\begin{equation}
    H(\s) = \sum_{(i, j) \in E} J_{ij} \s_i \s_j - \sum_{i = 1}^N h_i \s_i
\end{equation}
can be encoded in the same way by treating the local field terms as $p$-spin interactions with $p = 1$.

The generic model containing $p$-spin interactions has ${N \choose p} \sim N^p$
independent couplings. Thus, it is clear that its encoding in a SPIM via SPE
requires an SLM with that many pixels. As for the $p=2$ case, the technique
will have better scaling properties in the case of models with sparse tensors
$J$, as only the non-vanishing elements are allocated.

\section{Space and time complexity}

The two main resources that come into play in the context of SPIMs are the number
of SLM pixels and the number of time steps\footnote{Here, we consider
as a single time step one SLM/CMOS cycle. In principle, loading an array of $O(n)$ elements onto the SLM
requires at least $O(n)$ time steps due to the memory copy operation. However, in the context of SPIMs, the memory copy
operation is orders of magnitude faster than the SLM rise/fall time, which is of the order of milliseconds.
The computation time is therfore dominated by the SLM rise/fall time, hence
we consider the memory copy operation as instantaneous.}
required to determine the energy. We refer to these as the space and time
complexity, respectively. In this section we discuss the space and time
complexity of our method, and compare it to existing methods.

At the time of writing, the only available single-SLM methods for encoding
arbitrary Ising models on SPIMs are based on \textit{time-division multiplexing} (TDM)~\cite{wang_general_2024}
and \textit{wavelength-division multiplexing} (WDM)~\cite{luo_wavelength-division_2023}. Both methods decompose the couplings
matrix into a sum of rank-1 matrices and encode these simpler model separately,
either in different time steps or in different wavelengths. Moreover, in both
these methods, the SLM pixels are mapped to individual spins.
In contrast, our method \textit{Spin-product-encoding} (SPE), doesn't use any matrix decomposition,
encodes only non-zero couplings, maps the SLM pixels to spin products, and
requires always two time steps to determine the energy.
The space and time complexity of the three methods are summarized in the table
below.

\begin{center}
    \begin{tabular}{| l | c | c |}
        \hline
        & \textbf{Space complexity} & \textbf{Time complexity} \\
        \hline
        TDM~\cite{wang_general_2024} & $O(N)$ & $O(\rank(J))$ \\
        WDM~\cite{luo_wavelength-division_2023} & $O(N^2)$ & $O(1)$ \\
        SPE & $O(|E|)$ & $O(1)$ \\
        \hline
    \end{tabular}
\end{center}

TDM has better space complexity but suffers from a time complexity linear in the
rank of the couplings matrix. Both WDM and SPE offer constant time complexity.
The space complexity of SPE is $O(|E|)$, as opposed to $O(N^2)$ for WDM. Hence, both
methods become equivalent for a fully-connected graph, but SPE is preferable for sparse graphs.

%% file: schematic.tex
\begingroup%
  \makeatletter%
  \providecommand\color[2][]{%
    \errmessage{(Inkscape) Color is used for the text in Inkscape, but the package 'color.sty' is not loaded}%
    \renewcommand\color[2][]{}%
  }%
  \providecommand\transparent[1]{%
    \errmessage{(Inkscape) Transparency is used (non-zero) for the text in Inkscape, but the package 'transparent.sty' is not loaded}%
    \renewcommand\transparent[1]{}%
  }%
  \providecommand\rotatebox[2]{#2}%
  \newcommand*\fsize{\dimexpr\f@size pt\relax}%
  \newcommand*\lineheight[1]{\fontsize{\fsize}{#1\fsize}\selectfont}%
  \ifx\svgwidth\undefined%
    \setlength{\unitlength}{451.22937012bp}%
    \ifx\svgscale\undefined%
      \relax%
    \else%
      \setlength{\unitlength}{\unitlength * \real{\svgscale}}%
    \fi%
  \else%
    \setlength{\unitlength}{\svgwidth}%
  \fi%
  \global\let\svgwidth\undefined%
  \global\let\svgscale\undefined%
  \makeatother%
  \begin{picture}(1,0.48675884)%
    \lineheight{1}%
    \setlength\tabcolsep{0pt}%
    \put(0,0){\includegraphics[width=\unitlength,page=1]{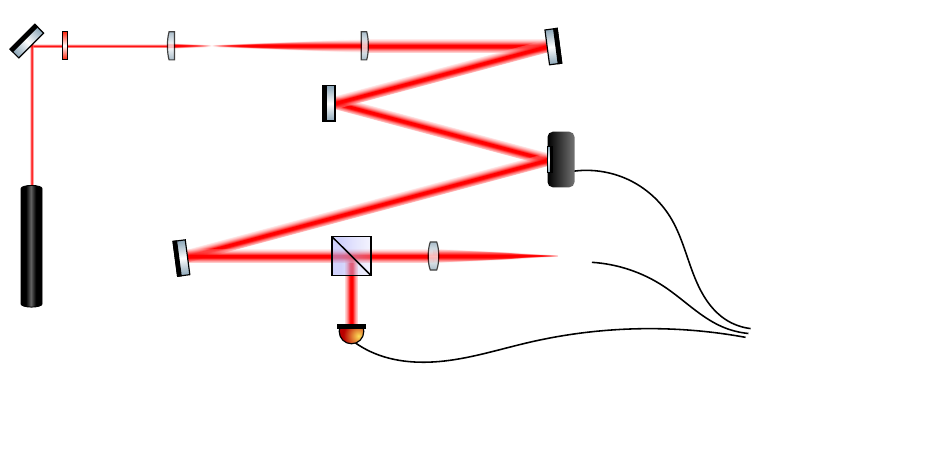}}%
    \put(0.052,0.21602693){\color[rgb]{0,0,0}\makebox(0,0)[lt]{\lineheight{1.25}\smash{\begin{tabular}[t]{l}Laser\end{tabular}}}}%
    \put(0.61756965,0.3113269){\color[rgb]{0,0,0}\makebox(0,0)[lt]{\lineheight{1.25}\smash{\begin{tabular}[t]{l}SLM\end{tabular}}}}%
    \put(0.64180823,0.20973585){\color[rgb]{0,0,0}\makebox(0,0)[lt]{\lineheight{1.25}\smash{\begin{tabular}[t]{l}CMOS\end{tabular}}}}%
    \put(0.13426574,0.12420995){\color[rgb]{0,0,0}\makebox(0,0)[lt]{\lineheight{1.25}\smash{\begin{tabular}[t]{l}Power meter\end{tabular}}}}%
    \put(0.30179256,0.17761194){\color[rgb]{0,0,0}\makebox(0,0)[lt]{\lineheight{1.25}\smash{\begin{tabular}[t]{l}BS\end{tabular}}}}%
    \put(0,0){\includegraphics[width=\unitlength,page=2]{schematic__pdf.pdf}}%
    \put(0.71260071,0.44691968){\color[rgb]{0,0,0}\makebox(0,0)[lt]{\lineheight{1.25}\smash{\begin{tabular}[t]{l}SLM phase mask\end{tabular}}}}%
    \put(0.86690526,0.27701662){\color[rgb]{0,0,0}\makebox(0,0)[lt]{\lineheight{1.25}\smash{\begin{tabular}[t]{l}PC\end{tabular}}}}%
    \put(0,0){\includegraphics[width=\unitlength,page=3]{schematic__pdf.pdf}}%
    \put(0.10299291,0.38649595){\color[rgb]{0,0,0}\makebox(0,0)[lt]{\lineheight{1.25}\smash{\begin{tabular}[t]{l}Polarizer\end{tabular}}}}%
    \put(0.16727442,0.46591199){\color[rgb]{0,0,0}\makebox(0,0)[lt]{\lineheight{1.25}\smash{\begin{tabular}[t]{l}$L_1$\end{tabular}}}}%
    \put(0.0601902,0.38772516){\color[rgb]{0.10196078,0.10196078,0.10196078}\makebox(0,0)[lt]{\lineheight{1.25}\smash{\begin{tabular}[t]{l}$\frac{\lambda}{2}$\end{tabular}}}}%
    \put(0.37301803,0.4658116){\color[rgb]{0,0,0}\makebox(0,0)[lt]{\lineheight{1.25}\smash{\begin{tabular}[t]{l}$L_2$\end{tabular}}}}%
    \put(0.44617355,0.24100176){\color[rgb]{0,0,0}\makebox(0,0)[lt]{\lineheight{1.25}\smash{\begin{tabular}[t]{l}$L_3$\end{tabular}}}}%
    \put(0,0){\includegraphics[width=\unitlength,page=4]{schematic__pdf.pdf}}%
    \put(0.46867826,0.03221998){\color[rgb]{0.10196078,0.10196078,0.10196078}\makebox(0,0)[lt]{\lineheight{1.25}\smash{\begin{tabular}[t]{l}(a)\end{tabular}}}}%
  \end{picture}%
\endgroup%

%% file: energy_scatter.tex
\begingroup
  \makeatletter
  \providecommand\color[2][]{%
    \GenericError{(gnuplot) \space\space\space\@spaces}{%
      Package color not loaded in conjunction with
      terminal option `colourtext'%
    }{See the gnuplot documentation for explanation.%
    }{Either use 'blacktext' in gnuplot or load the package
      color.sty in LaTeX.}%
    \renewcommand\color[2][]{}%
  }%
  \providecommand\includegraphics[2][]{%
    \GenericError{(gnuplot) \space\space\space\@spaces}{%
      Package graphicx or graphics not loaded%
    }{See the gnuplot documentation for explanation.%
    }{The gnuplot epslatex terminal needs graphicx.sty or graphics.sty.}%
    \renewcommand\includegraphics[2][]{}%
  }%
  \providecommand\rotatebox[2]{#2}%
  \@ifundefined{ifGPcolor}{%
    \newif\ifGPcolor
    \GPcolortrue
  }{}%
  \@ifundefined{ifGPblacktext}{%
    \newif\ifGPblacktext
    \GPblacktexttrue
  }{}%
  \let\gplgaddtomacro\g@addto@macro
  \gdef\gplbacktext{}%
  \gdef\gplfronttext{}%
  \makeatother
  \ifGPblacktext
    \def\colorrgb#1{}%
    \def\colorgray#1{}%
  \else
    \ifGPcolor
      \def\colorrgb#1{\color[rgb]{#1}}%
      \def\colorgray#1{\color[gray]{#1}}%
      \expandafter\def\csname LTw\endcsname{\color{white}}%
      \expandafter\def\csname LTb\endcsname{\color{black}}%
      \expandafter\def\csname LTa\endcsname{\color{black}}%
      \expandafter\def\csname LT0\endcsname{\color[rgb]{1,0,0}}%
      \expandafter\def\csname LT1\endcsname{\color[rgb]{0,1,0}}%
      \expandafter\def\csname LT2\endcsname{\color[rgb]{0,0,1}}%
      \expandafter\def\csname LT3\endcsname{\color[rgb]{1,0,1}}%
      \expandafter\def\csname LT4\endcsname{\color[rgb]{0,1,1}}%
      \expandafter\def\csname LT5\endcsname{\color[rgb]{1,1,0}}%
      \expandafter\def\csname LT6\endcsname{\color[rgb]{0,0,0}}%
      \expandafter\def\csname LT7\endcsname{\color[rgb]{1,0.3,0}}%
      \expandafter\def\csname LT8\endcsname{\color[rgb]{0.5,0.5,0.5}}%
    \else
      \def\colorrgb#1{\color{black}}%
      \def\colorgray#1{\color[gray]{#1}}%
      \expandafter\def\csname LTw\endcsname{\color{white}}%
      \expandafter\def\csname LTb\endcsname{\color{black}}%
      \expandafter\def\csname LTa\endcsname{\color{black}}%
      \expandafter\def\csname LT0\endcsname{\color{black}}%
      \expandafter\def\csname LT1\endcsname{\color{black}}%
      \expandafter\def\csname LT2\endcsname{\color{black}}%
      \expandafter\def\csname LT3\endcsname{\color{black}}%
      \expandafter\def\csname LT4\endcsname{\color{black}}%
      \expandafter\def\csname LT5\endcsname{\color{black}}%
      \expandafter\def\csname LT6\endcsname{\color{black}}%
      \expandafter\def\csname LT7\endcsname{\color{black}}%
      \expandafter\def\csname LT8\endcsname{\color{black}}%
    \fi
  \fi
    \setlength{\unitlength}{0.0500bp}%
    \ifx\gptboxheight\undefined%
      \newlength{\gptboxheight}%
      \newlength{\gptboxwidth}%
      \newsavebox{\gptboxtext}%
    \fi%
    \setlength{\fboxrule}{0.5pt}%
    \setlength{\fboxsep}{1pt}%
    \definecolor{tbcol}{rgb}{1,1,1}%
\begin{picture}(7200.00,3600.00)%
    \gplgaddtomacro\gplbacktext{%
      \csname LTb\endcsname
      \put(826,744){\makebox(0,0)[r]{\strut{}$-400$}}%
      \csname LTb\endcsname
      \put(826,1265){\makebox(0,0)[r]{\strut{}$-300$}}%
      \csname LTb\endcsname
      \put(826,1787){\makebox(0,0)[r]{\strut{}$-200$}}%
      \csname LTb\endcsname
      \put(826,2308){\makebox(0,0)[r]{\strut{}$-100$}}%
      \csname LTb\endcsname
      \put(826,2830){\makebox(0,0)[r]{\strut{}$0$}}%
      \csname LTb\endcsname
      \put(826,3351){\makebox(0,0)[r]{\strut{}$100$}}%
      \csname LTb\endcsname
      \put(1187,291){\makebox(0,0){\strut{}$-400$}}%
      \csname LTb\endcsname
      \put(1694,291){\makebox(0,0){\strut{}$-300$}}%
      \csname LTb\endcsname
      \put(2202,291){\makebox(0,0){\strut{}$-200$}}%
      \csname LTb\endcsname
      \put(2709,291){\makebox(0,0){\strut{}$-100$}}%
      \csname LTb\endcsname
      \put(3217,291){\makebox(0,0){\strut{}$0$}}%
      \csname LTb\endcsname
      \put(3724,291){\makebox(0,0){\strut{}$100$}}%
      \csname LTb\endcsname
      \put(1227,3205){\makebox(0,0){\large (c)}}%
    }%
    \gplgaddtomacro\gplfronttext{%
      \csname LTb\endcsname
      \put(223,1995){\rotatebox{-270.00}{\makebox(0,0){\strut{}$H_{\text{exp}}$}}}%
    }%
    \gplgaddtomacro\gplbacktext{%
      \csname LTb\endcsname
      \put(4245,291){\makebox(0,0){\strut{}$-40$}}%
      \csname LTb\endcsname
      \put(4981,291){\makebox(0,0){\strut{}$-20$}}%
      \csname LTb\endcsname
      \put(5717,291){\makebox(0,0){\strut{}$0$}}%
      \csname LTb\endcsname
      \put(6452,291){\makebox(0,0){\strut{}$20$}}%
      \csname LTb\endcsname
      \put(6927,861){\makebox(0,0)[l]{\strut{}$-40$}}%
      \csname LTb\endcsname
      \put(6927,1617){\makebox(0,0)[l]{\strut{}$-20$}}%
      \csname LTb\endcsname
      \put(6927,2373){\makebox(0,0)[l]{\strut{}$0$}}%
      \csname LTb\endcsname
      \put(6927,3130){\makebox(0,0)[l]{\strut{}$20$}}%
      \csname LTb\endcsname
      \put(4171,3205){\makebox(0,0){\large (d)}}%
      \csname LTb\endcsname
      \put(3733,0){\makebox(0,0){\strut{}$H_{\text{theory}}$}}%
    }%
    \gplgaddtomacro\gplfronttext{%
    }%
    \gplbacktext
    \put(0,0){\includegraphics[width={360.00bp},height={180.00bp}]{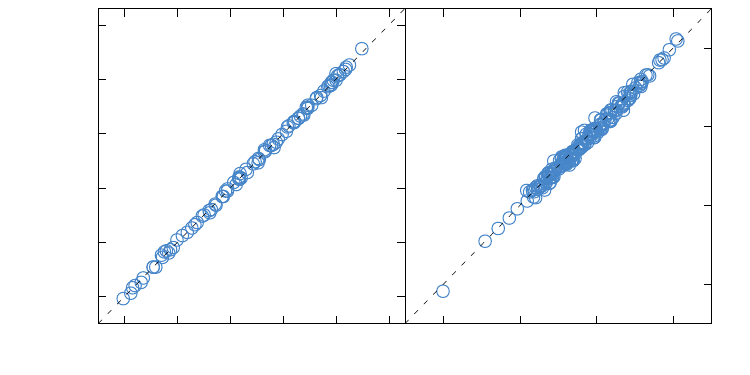}}%
    \gplfronttext
  \end{picture}%
\endgroup

%% file: graph_part_sim_and_spim.tex
\begingroup
  \makeatletter
  \providecommand\color[2][]{%
    \GenericError{(gnuplot) \space\space\space\@spaces}{%
      Package color not loaded in conjunction with
      terminal option `colourtext'%
    }{See the gnuplot documentation for explanation.%
    }{Either use 'blacktext' in gnuplot or load the package
      color.sty in LaTeX.}%
    \renewcommand\color[2][]{}%
  }%
  \providecommand\includegraphics[2][]{%
    \GenericError{(gnuplot) \space\space\space\@spaces}{%
      Package graphicx or graphics not loaded%
    }{See the gnuplot documentation for explanation.%
    }{The gnuplot epslatex terminal needs graphicx.sty or graphics.sty.}%
    \renewcommand\includegraphics[2][]{}%
  }%
  \providecommand\rotatebox[2]{#2}%
  \@ifundefined{ifGPcolor}{%
    \newif\ifGPcolor
    \GPcolortrue
  }{}%
  \@ifundefined{ifGPblacktext}{%
    \newif\ifGPblacktext
    \GPblacktexttrue
  }{}%
  \let\gplgaddtomacro\g@addto@macro
  \gdef\gplbacktext{}%
  \gdef\gplfronttext{}%
  \makeatother
  \ifGPblacktext
    \def\colorrgb#1{}%
    \def\colorgray#1{}%
  \else
    \ifGPcolor
      \def\colorrgb#1{\color[rgb]{#1}}%
      \def\colorgray#1{\color[gray]{#1}}%
      \expandafter\def\csname LTw\endcsname{\color{white}}%
      \expandafter\def\csname LTb\endcsname{\color{black}}%
      \expandafter\def\csname LTa\endcsname{\color{black}}%
      \expandafter\def\csname LT0\endcsname{\color[rgb]{1,0,0}}%
      \expandafter\def\csname LT1\endcsname{\color[rgb]{0,1,0}}%
      \expandafter\def\csname LT2\endcsname{\color[rgb]{0,0,1}}%
      \expandafter\def\csname LT3\endcsname{\color[rgb]{1,0,1}}%
      \expandafter\def\csname LT4\endcsname{\color[rgb]{0,1,1}}%
      \expandafter\def\csname LT5\endcsname{\color[rgb]{1,1,0}}%
      \expandafter\def\csname LT6\endcsname{\color[rgb]{0,0,0}}%
      \expandafter\def\csname LT7\endcsname{\color[rgb]{1,0.3,0}}%
      \expandafter\def\csname LT8\endcsname{\color[rgb]{0.5,0.5,0.5}}%
    \else
      \def\colorrgb#1{\color{black}}%
      \def\colorgray#1{\color[gray]{#1}}%
      \expandafter\def\csname LTw\endcsname{\color{white}}%
      \expandafter\def\csname LTb\endcsname{\color{black}}%
      \expandafter\def\csname LTa\endcsname{\color{black}}%
      \expandafter\def\csname LT0\endcsname{\color{black}}%
      \expandafter\def\csname LT1\endcsname{\color{black}}%
      \expandafter\def\csname LT2\endcsname{\color{black}}%
      \expandafter\def\csname LT3\endcsname{\color{black}}%
      \expandafter\def\csname LT4\endcsname{\color{black}}%
      \expandafter\def\csname LT5\endcsname{\color{black}}%
      \expandafter\def\csname LT6\endcsname{\color{black}}%
      \expandafter\def\csname LT7\endcsname{\color{black}}%
      \expandafter\def\csname LT8\endcsname{\color{black}}%
    \fi
  \fi
    \setlength{\unitlength}{0.0500bp}%
    \ifx\gptboxheight\undefined%
      \newlength{\gptboxheight}%
      \newlength{\gptboxwidth}%
      \newsavebox{\gptboxtext}%
    \fi%
    \setlength{\fboxrule}{0.5pt}%
    \setlength{\fboxsep}{1pt}%
    \definecolor{tbcol}{rgb}{1,1,1}%
\begin{picture}(7200.00,5760.00)%
    \gplgaddtomacro\gplbacktext{%
      \csname LTb\endcsname
      \put(252,3819){\makebox(0,0)[r]{\strut{}$0$}}%
      \csname LTb\endcsname
      \put(252,4137){\makebox(0,0)[r]{\strut{}$50$}}%
      \csname LTb\endcsname
      \put(252,4454){\makebox(0,0)[r]{\strut{}$100$}}%
      \csname LTb\endcsname
      \put(252,4772){\makebox(0,0)[r]{\strut{}$150$}}%
      \csname LTb\endcsname
      \put(252,5090){\makebox(0,0)[r]{\strut{}$200$}}%
      \csname LTb\endcsname
      \put(359,3500){\makebox(0,0){\strut{}}}%
      \csname LTb\endcsname
      \put(1112,3500){\makebox(0,0){\strut{}}}%
      \csname LTb\endcsname
      \put(1866,3500){\makebox(0,0){\strut{}}}%
      \csname LTb\endcsname
      \put(2620,3500){\makebox(0,0){\strut{}}}%
      \csname LTb\endcsname
      \put(3373,3500){\makebox(0,0){\strut{}}}%
      \csname LTb\endcsname
      \put(1472,4963){\makebox(0,0)[l]{\strut{}$H$}}%
      \csname LTb\endcsname
      \put(718,4582){\makebox(0,0)[l]{\strut{}$H_b$}}%
      \csname LTb\endcsname
      \put(1112,4073){\makebox(0,0)[l]{\strut{}$H_a$}}%
      \csname LTb\endcsname
      \put(3396,4994){\makebox(0,0){\large (a)}}%
    }%
    \gplgaddtomacro\gplfronttext{%
      \csname LTb\endcsname
      \put(2046,5472){\makebox(0,0){\strut{}SPIM Simulation}}%
    }%
    \gplgaddtomacro\gplbacktext{%
      \csname LTb\endcsname
      \put(252,2176){\makebox(0,0)[r]{\strut{}$-0.2$}}%
      \csname LTb\endcsname
      \put(252,2537){\makebox(0,0)[r]{\strut{}$-0.1$}}%
      \csname LTb\endcsname
      \put(252,2898){\makebox(0,0)[r]{\strut{}$0.0$}}%
      \csname LTb\endcsname
      \put(252,3259){\makebox(0,0)[r]{\strut{}$0.1$}}%
      \csname LTb\endcsname
      \put(252,3620){\makebox(0,0)[r]{\strut{}$0.2$}}%
      \csname LTb\endcsname
      \put(359,1912){\makebox(0,0){\strut{}}}%
      \csname LTb\endcsname
      \put(1112,1912){\makebox(0,0){\strut{}}}%
      \csname LTb\endcsname
      \put(1866,1912){\makebox(0,0){\strut{}}}%
      \csname LTb\endcsname
      \put(2620,1912){\makebox(0,0){\strut{}}}%
      \csname LTb\endcsname
      \put(3373,1912){\makebox(0,0){\strut{}}}%
      \csname LTb\endcsname
      \put(2093,3151){\makebox(0,0)[l]{\strut{}M}}%
      \csname LTb\endcsname
      \put(585,3440){\makebox(0,0)[l]{\strut{}$M^*_{\text{metis}}=0$}}%
      \csname LTb\endcsname
      \put(2706,2393){\makebox(0,0)[l]{\strut{}$M^*_{\text{sim}}=0$}}%
      \csname LTb\endcsname
      \put(3396,3406){\makebox(0,0){\large (b)}}%
    }%
    \gplgaddtomacro\gplfronttext{%
    }%
    \gplgaddtomacro\gplbacktext{%
      \csname LTb\endcsname
      \put(252,516){\makebox(0,0)[r]{\strut{}$50$}}%
      \csname LTb\endcsname
      \put(252,1238){\makebox(0,0)[r]{\strut{}$100$}}%
      \csname LTb\endcsname
      \put(252,1960){\makebox(0,0)[r]{\strut{}$150$}}%
      \csname LTb\endcsname
      \put(359,324){\makebox(0,0){\strut{}$10^{0}$}}%
      \csname LTb\endcsname
      \put(1112,324){\makebox(0,0){\strut{}$10^{1}$}}%
      \csname LTb\endcsname
      \put(1866,324){\makebox(0,0){\strut{}$10^{2}$}}%
      \csname LTb\endcsname
      \put(2620,324){\makebox(0,0){\strut{}$10^{3}$}}%
      \csname LTb\endcsname
      \put(3373,324){\makebox(0,0){\strut{}$10^{4}$}}%
      \csname LTb\endcsname
      \put(1866,1815){\makebox(0,0)[l]{\strut{}C}}%
      \csname LTb\endcsname
      \put(1639,805){\makebox(0,0)[l]{\strut{}$C^*_{\text{metis}}=60$}}%
      \csname LTb\endcsname
      \put(2679,1382){\makebox(0,0)[l]{\strut{}$C^*_{\text{sim}}=61$}}%
      \csname LTb\endcsname
      \put(3396,1818){\makebox(0,0){\large (c)}}%
    }%
    \gplgaddtomacro\gplfronttext{%
    }%
    \gplgaddtomacro\gplbacktext{%
      \csname LTb\endcsname
      \put(3626,3819){\makebox(0,0)[r]{\strut{}}}%
      \csname LTb\endcsname
      \put(3626,4137){\makebox(0,0)[r]{\strut{}}}%
      \csname LTb\endcsname
      \put(3626,4454){\makebox(0,0)[r]{\strut{}}}%
      \csname LTb\endcsname
      \put(3626,4772){\makebox(0,0)[r]{\strut{}}}%
      \csname LTb\endcsname
      \put(3626,5090){\makebox(0,0)[r]{\strut{}}}%
      \csname LTb\endcsname
      \put(3733,3500){\makebox(0,0){\strut{}}}%
      \csname LTb\endcsname
      \put(4487,3500){\makebox(0,0){\strut{}}}%
      \csname LTb\endcsname
      \put(5241,3500){\makebox(0,0){\strut{}}}%
      \csname LTb\endcsname
      \put(5994,3500){\makebox(0,0){\strut{}}}%
      \csname LTb\endcsname
      \put(6748,3500){\makebox(0,0){\strut{}}}%
      \csname LTb\endcsname
      \put(4487,5090){\makebox(0,0)[l]{\strut{}$H$}}%
      \csname LTb\endcsname
      \put(3960,4772){\makebox(0,0)[l]{\strut{}$H_b$}}%
      \csname LTb\endcsname
      \put(3960,4073){\makebox(0,0)[l]{\strut{}$H_a$}}%
      \csname LTb\endcsname
      \put(6770,4994){\makebox(0,0){\large (d)}}%
    }%
    \gplgaddtomacro\gplfronttext{%
      \csname LTb\endcsname
      \put(5420,5472){\makebox(0,0){\strut{}SPIM Experiment}}%
    }%
    \gplgaddtomacro\gplbacktext{%
      \csname LTb\endcsname
      \put(3626,2176){\makebox(0,0)[r]{\strut{}}}%
      \csname LTb\endcsname
      \put(3626,2537){\makebox(0,0)[r]{\strut{}}}%
      \csname LTb\endcsname
      \put(3626,2898){\makebox(0,0)[r]{\strut{}}}%
      \csname LTb\endcsname
      \put(3626,3259){\makebox(0,0)[r]{\strut{}}}%
      \csname LTb\endcsname
      \put(3626,3620){\makebox(0,0)[r]{\strut{}}}%
      \csname LTb\endcsname
      \put(3733,1912){\makebox(0,0){\strut{}}}%
      \csname LTb\endcsname
      \put(4487,1912){\makebox(0,0){\strut{}}}%
      \csname LTb\endcsname
      \put(5241,1912){\makebox(0,0){\strut{}}}%
      \csname LTb\endcsname
      \put(5994,1912){\makebox(0,0){\strut{}}}%
      \csname LTb\endcsname
      \put(6748,1912){\makebox(0,0){\strut{}}}%
      \csname LTb\endcsname
      \put(5467,3151){\makebox(0,0)[l]{\strut{}M}}%
      \csname LTb\endcsname
      \put(3793,3440){\makebox(0,0)[l]{\strut{}$M^*_{\text{metis}}=0$}}%
      \csname LTb\endcsname
      \put(6080,2393){\makebox(0,0)[l]{\strut{}$M^*_{\text{exp}}=0$}}%
      \csname LTb\endcsname
      \put(6770,3406){\makebox(0,0){\large (e)}}%
    }%
    \gplgaddtomacro\gplfronttext{%
    }%
    \gplgaddtomacro\gplbacktext{%
      \csname LTb\endcsname
      \put(3626,516){\makebox(0,0)[r]{\strut{}}}%
      \csname LTb\endcsname
      \put(3626,1238){\makebox(0,0)[r]{\strut{}}}%
      \csname LTb\endcsname
      \put(3626,1960){\makebox(0,0)[r]{\strut{}}}%
      \csname LTb\endcsname
      \put(3733,324){\makebox(0,0){\strut{}$10^{0}$}}%
      \csname LTb\endcsname
      \put(4487,324){\makebox(0,0){\strut{}$10^{1}$}}%
      \csname LTb\endcsname
      \put(5241,324){\makebox(0,0){\strut{}$10^{2}$}}%
      \csname LTb\endcsname
      \put(5994,324){\makebox(0,0){\strut{}$10^{3}$}}%
      \csname LTb\endcsname
      \put(6748,324){\makebox(0,0){\strut{}$10^{4}$}}%
      \csname LTb\endcsname
      \put(5241,1671){\makebox(0,0)[l]{\strut{}C}}%
      \csname LTb\endcsname
      \put(5014,805){\makebox(0,0)[l]{\strut{}$C^*_{\text{metis}}=60$}}%
      \csname LTb\endcsname
      \put(6054,1382){\makebox(0,0)[l]{\strut{}$C^*_{\text{exp}}=58$}}%
      \csname LTb\endcsname
      \put(6770,1818){\makebox(0,0){\large (f)}}%
      \csname LTb\endcsname
      \put(3590,80){\makebox(0,0){\strut{}Iteration}}%
    }%
    \gplgaddtomacro\gplfronttext{%
    }%
    \gplgaddtomacro\gplbacktext{%
      \csname LTb\endcsname
      \put(996,1498){\makebox(0,0){\tiny SA schedules}}%
    }%
    \gplgaddtomacro\gplfronttext{%
    }%
    \gplgaddtomacro\gplbacktext{%
    }%
    \gplgaddtomacro\gplfronttext{%
    }%
    \gplbacktext
    \put(0,0){\includegraphics[width={360.00bp},height={288.00bp}]{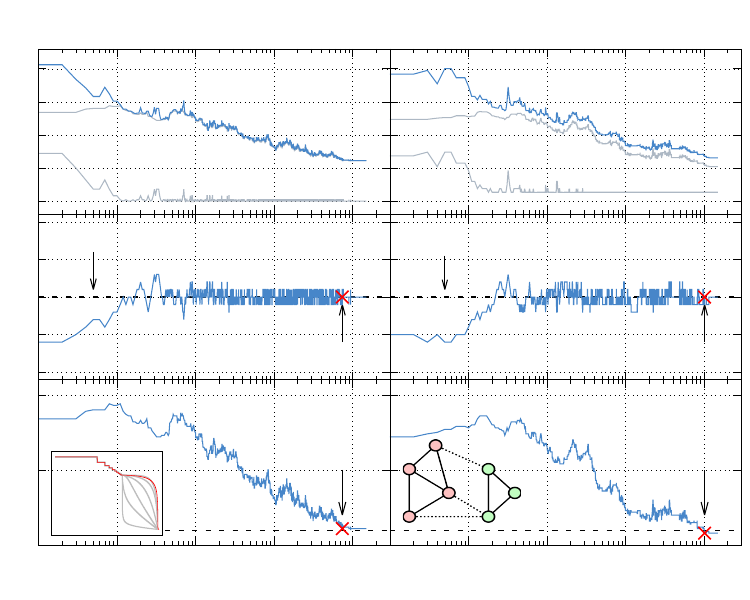}}%
    \gplfronttext
  \end{picture}%
\endgroup

%% file: stats-gpp-wgpp.tex
\begingroup
  \makeatletter
  \providecommand\color[2][]{%
    \GenericError{(gnuplot) \space\space\space\@spaces}{%
      Package color not loaded in conjunction with
      terminal option `colourtext'%
    }{See the gnuplot documentation for explanation.%
    }{Either use 'blacktext' in gnuplot or load the package
      color.sty in LaTeX.}%
    \renewcommand\color[2][]{}%
  }%
  \providecommand\includegraphics[2][]{%
    \GenericError{(gnuplot) \space\space\space\@spaces}{%
      Package graphicx or graphics not loaded%
    }{See the gnuplot documentation for explanation.%
    }{The gnuplot epslatex terminal needs graphicx.sty or graphics.sty.}%
    \renewcommand\includegraphics[2][]{}%
  }%
  \providecommand\rotatebox[2]{#2}%
  \@ifundefined{ifGPcolor}{%
    \newif\ifGPcolor
    \GPcolortrue
  }{}%
  \@ifundefined{ifGPblacktext}{%
    \newif\ifGPblacktext
    \GPblacktexttrue
  }{}%
  \let\gplgaddtomacro\g@addto@macro
  \gdef\gplbacktext{}%
  \gdef\gplfronttext{}%
  \makeatother
  \ifGPblacktext
    \def\colorrgb#1{}%
    \def\colorgray#1{}%
  \else
    \ifGPcolor
      \def\colorrgb#1{\color[rgb]{#1}}%
      \def\colorgray#1{\color[gray]{#1}}%
      \expandafter\def\csname LTw\endcsname{\color{white}}%
      \expandafter\def\csname LTb\endcsname{\color{black}}%
      \expandafter\def\csname LTa\endcsname{\color{black}}%
      \expandafter\def\csname LT0\endcsname{\color[rgb]{1,0,0}}%
      \expandafter\def\csname LT1\endcsname{\color[rgb]{0,1,0}}%
      \expandafter\def\csname LT2\endcsname{\color[rgb]{0,0,1}}%
      \expandafter\def\csname LT3\endcsname{\color[rgb]{1,0,1}}%
      \expandafter\def\csname LT4\endcsname{\color[rgb]{0,1,1}}%
      \expandafter\def\csname LT5\endcsname{\color[rgb]{1,1,0}}%
      \expandafter\def\csname LT6\endcsname{\color[rgb]{0,0,0}}%
      \expandafter\def\csname LT7\endcsname{\color[rgb]{1,0.3,0}}%
      \expandafter\def\csname LT8\endcsname{\color[rgb]{0.5,0.5,0.5}}%
    \else
      \def\colorrgb#1{\color{black}}%
      \def\colorgray#1{\color[gray]{#1}}%
      \expandafter\def\csname LTw\endcsname{\color{white}}%
      \expandafter\def\csname LTb\endcsname{\color{black}}%
      \expandafter\def\csname LTa\endcsname{\color{black}}%
      \expandafter\def\csname LT0\endcsname{\color{black}}%
      \expandafter\def\csname LT1\endcsname{\color{black}}%
      \expandafter\def\csname LT2\endcsname{\color{black}}%
      \expandafter\def\csname LT3\endcsname{\color{black}}%
      \expandafter\def\csname LT4\endcsname{\color{black}}%
      \expandafter\def\csname LT5\endcsname{\color{black}}%
      \expandafter\def\csname LT6\endcsname{\color{black}}%
      \expandafter\def\csname LT7\endcsname{\color{black}}%
      \expandafter\def\csname LT8\endcsname{\color{black}}%
    \fi
  \fi
    \setlength{\unitlength}{0.0500bp}%
    \ifx\gptboxheight\undefined%
      \newlength{\gptboxheight}%
      \newlength{\gptboxwidth}%
      \newsavebox{\gptboxtext}%
    \fi%
    \setlength{\fboxrule}{0.5pt}%
    \setlength{\fboxsep}{1pt}%
    \definecolor{tbcol}{rgb}{1,1,1}%
\begin{picture}(7200.00,5760.00)%
    \gplgaddtomacro\gplbacktext{%
      \csname LTb\endcsname
      \put(395,3466){\makebox(0,0)[r]{\strut{}$80$}}%
      \csname LTb\endcsname
      \put(395,4109){\makebox(0,0)[r]{\strut{}$120$}}%
      \csname LTb\endcsname
      \put(395,4752){\makebox(0,0)[r]{\strut{}$160$}}%
      \csname LTb\endcsname
      \put(395,5395){\makebox(0,0)[r]{\strut{}$200$}}%
      \csname LTb\endcsname
      \put(502,2792){\makebox(0,0){\strut{}}}%
      \csname LTb\endcsname
      \put(1168,2792){\makebox(0,0){\strut{}}}%
      \csname LTb\endcsname
      \put(1833,2792){\makebox(0,0){\strut{}}}%
      \csname LTb\endcsname
      \put(2499,2792){\makebox(0,0){\strut{}}}%
      \csname LTb\endcsname
      \put(3164,2792){\makebox(0,0){\strut{}}}%
      \csname LTb\endcsname
      \put(1568,4752){\makebox(0,0)[l]{\strut{}$H$}}%
      \csname LTb\endcsname
      \put(3184,5154){\makebox(0,0){\large (a)}}%
    }%
    \gplgaddtomacro\gplfronttext{%
      \csname LTb\endcsname
      \put(1992,5683){\makebox(0,0){\strut{}Unweighted graph partitioning}}%
    }%
    \gplgaddtomacro\gplbacktext{%
      \csname LTb\endcsname
      \put(395,1141){\makebox(0,0)[r]{\strut{}$75$}}%
      \csname LTb\endcsname
      \put(395,1850){\makebox(0,0)[r]{\strut{}$100$}}%
      \csname LTb\endcsname
      \put(395,2559){\makebox(0,0)[r]{\strut{}$125$}}%
      \csname LTb\endcsname
      \put(502,382){\makebox(0,0){\strut{}$10^{0}$}}%
      \csname LTb\endcsname
      \put(1168,382){\makebox(0,0){\strut{}$10^{1}$}}%
      \csname LTb\endcsname
      \put(1833,382){\makebox(0,0){\strut{}$10^{2}$}}%
      \csname LTb\endcsname
      \put(2499,382){\makebox(0,0){\strut{}$10^{3}$}}%
      \csname LTb\endcsname
      \put(3164,382){\makebox(0,0){\strut{}$10^{4}$}}%
      \csname LTb\endcsname
      \put(2034,2559){\makebox(0,0)[l]{\strut{}C}}%
      \csname LTb\endcsname
      \put(2499,829){\makebox(0,0){\scriptsize $C^*_{\text{metis}}=60$}}%
      \csname LTb\endcsname
      \put(2861,1992){\makebox(0,0){\scriptsize $C^*_{\text{exp}}=62.0 \pm 1.8$}}%
      \csname LTb\endcsname
      \put(3184,2743){\makebox(0,0){\large (b)}}%
    }%
    \gplgaddtomacro\gplfronttext{%
    }%
    \gplgaddtomacro\gplbacktext{%
      \csname LTb\endcsname
      \put(3482,2792){\makebox(0,0){\strut{}}}%
      \csname LTb\endcsname
      \put(4175,2792){\makebox(0,0){\strut{}}}%
      \csname LTb\endcsname
      \put(4867,2792){\makebox(0,0){\strut{}}}%
      \csname LTb\endcsname
      \put(5560,2792){\makebox(0,0){\strut{}}}%
      \csname LTb\endcsname
      \put(6253,2792){\makebox(0,0){\strut{}}}%
      \csname LTb\endcsname
      \put(6568,3286){\makebox(0,0)[l]{\strut{}$100$}}%
      \csname LTb\endcsname
      \put(6568,3888){\makebox(0,0)[l]{\strut{}$200$}}%
      \csname LTb\endcsname
      \put(6568,4491){\makebox(0,0)[l]{\strut{}$300$}}%
      \csname LTb\endcsname
      \put(6568,5094){\makebox(0,0)[l]{\strut{}$400$}}%
      \csname LTb\endcsname
      \put(4021,4491){\makebox(0,0)[l]{\strut{}$H$}}%
      \csname LTb\endcsname
      \put(6163,5154){\makebox(0,0){\large (c)}}%
    }%
    \gplgaddtomacro\gplfronttext{%
      \csname LTb\endcsname
      \put(4972,5683){\makebox(0,0){\strut{}Weighted graph partitioning}}%
    }%
    \gplgaddtomacro\gplbacktext{%
      \csname LTb\endcsname
      \put(3482,382){\makebox(0,0){\strut{}$10^{0}$}}%
      \csname LTb\endcsname
      \put(4175,382){\makebox(0,0){\strut{}$10^{1}$}}%
      \csname LTb\endcsname
      \put(4867,382){\makebox(0,0){\strut{}$10^{2}$}}%
      \csname LTb\endcsname
      \put(5560,382){\makebox(0,0){\strut{}$10^{3}$}}%
      \csname LTb\endcsname
      \put(6253,382){\makebox(0,0){\strut{}$10^{4}$}}%
      \csname LTb\endcsname
      \put(6568,574){\makebox(0,0)[l]{\strut{}$2 \cdot 10^{3}$}}%
      \csname LTb\endcsname
      \put(6568,1538){\makebox(0,0)[l]{\strut{}$4 \cdot 10^{3}$}}%
      \csname LTb\endcsname
      \put(6568,2502){\makebox(0,0)[l]{\strut{}$6 \cdot 10^{3}$}}%
      \csname LTb\endcsname
      \put(4867,2502){\makebox(0,0)[l]{\strut{}C}}%
      \csname LTb\endcsname
      \put(5560,959){\makebox(0,0){\scriptsize $C^*_{\text{metis}}=2539$}}%
      \csname LTb\endcsname
      \put(5797,1875){\makebox(0,0){\scriptsize $C^*_{\text{exp}}= 2442 \pm 157$}}%
      \csname LTb\endcsname
      \put(6163,2743){\makebox(0,0){\large (d)}}%
    }%
    \gplgaddtomacro\gplfronttext{%
    }%
    \gplgaddtomacro\gplbacktext{%
    }%
    \gplgaddtomacro\gplfronttext{%
    }%
    \gplgaddtomacro\gplbacktext{%
    }%
    \gplgaddtomacro\gplfronttext{%
    }%
    \gplgaddtomacro\gplbacktext{%
      \csname LTb\endcsname
      \put(4927,4341){\makebox(0,0)[r]{\strut{}}}%
      \csname LTb\endcsname
      \put(4927,4642){\makebox(0,0)[r]{\strut{}}}%
      \csname LTb\endcsname
      \put(4927,4942){\makebox(0,0)[r]{\strut{}}}%
      \csname LTb\endcsname
      \put(4927,5242){\makebox(0,0)[r]{\strut{}}}%
    }%
    \gplgaddtomacro\gplfronttext{%
    }%
    \gplgaddtomacro\gplbacktext{%
      \csname LTb\endcsname
      \put(3376,845){\makebox(0,0)[r]{\strut{}}}%
      \csname LTb\endcsname
      \put(3376,1207){\makebox(0,0)[r]{\strut{}}}%
      \csname LTb\endcsname
      \put(3376,1569){\makebox(0,0)[r]{\strut{}}}%
      \csname LTb\endcsname
      \put(3376,1931){\makebox(0,0)[r]{\strut{}}}%
      \csname LTb\endcsname
      \put(3590,80){\makebox(0,0){\strut{}Iteration}}%
    }%
    \gplgaddtomacro\gplfronttext{%
    }%
    \gplbacktext
    \put(0,0){\includegraphics[width={360.00bp},height={288.00bp}]{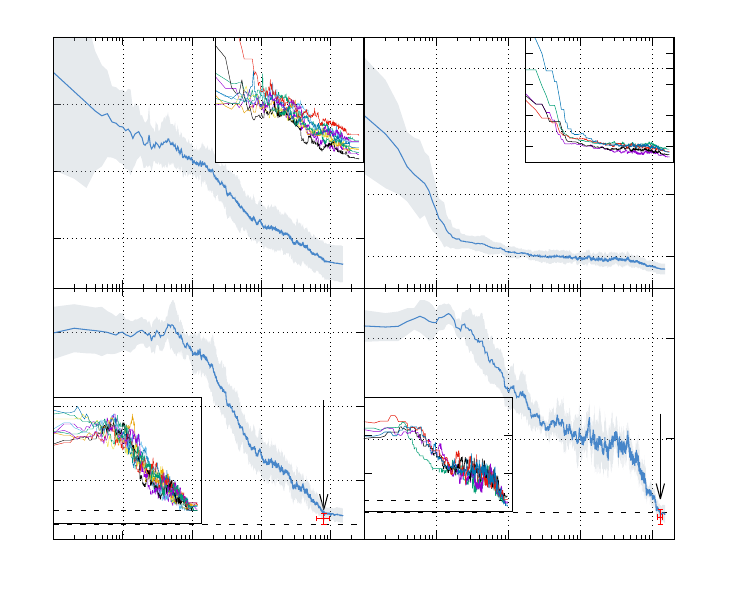}}%
    \gplfronttext
  \end{picture}%
\endgroup

%% file: solution_vs_density.tex
\begingroup
  \makeatletter
  \providecommand\color[2][]{%
    \GenericError{(gnuplot) \space\space\space\@spaces}{%
      Package color not loaded in conjunction with
      terminal option `colourtext'%
    }{See the gnuplot documentation for explanation.%
    }{Either use 'blacktext' in gnuplot or load the package
      color.sty in LaTeX.}%
    \renewcommand\color[2][]{}%
  }%
  \providecommand\includegraphics[2][]{%
    \GenericError{(gnuplot) \space\space\space\@spaces}{%
      Package graphicx or graphics not loaded%
    }{See the gnuplot documentation for explanation.%
    }{The gnuplot epslatex terminal needs graphicx.sty or graphics.sty.}%
    \renewcommand\includegraphics[2][]{}%
  }%
  \providecommand\rotatebox[2]{#2}%
  \@ifundefined{ifGPcolor}{%
    \newif\ifGPcolor
    \GPcolortrue
  }{}%
  \@ifundefined{ifGPblacktext}{%
    \newif\ifGPblacktext
    \GPblacktexttrue
  }{}%
  \let\gplgaddtomacro\g@addto@macro
  \gdef\gplbacktext{}%
  \gdef\gplfronttext{}%
  \makeatother
  \ifGPblacktext
    \def\colorrgb#1{}%
    \def\colorgray#1{}%
  \else
    \ifGPcolor
      \def\colorrgb#1{\color[rgb]{#1}}%
      \def\colorgray#1{\color[gray]{#1}}%
      \expandafter\def\csname LTw\endcsname{\color{white}}%
      \expandafter\def\csname LTb\endcsname{\color{black}}%
      \expandafter\def\csname LTa\endcsname{\color{black}}%
      \expandafter\def\csname LT0\endcsname{\color[rgb]{1,0,0}}%
      \expandafter\def\csname LT1\endcsname{\color[rgb]{0,1,0}}%
      \expandafter\def\csname LT2\endcsname{\color[rgb]{0,0,1}}%
      \expandafter\def\csname LT3\endcsname{\color[rgb]{1,0,1}}%
      \expandafter\def\csname LT4\endcsname{\color[rgb]{0,1,1}}%
      \expandafter\def\csname LT5\endcsname{\color[rgb]{1,1,0}}%
      \expandafter\def\csname LT6\endcsname{\color[rgb]{0,0,0}}%
      \expandafter\def\csname LT7\endcsname{\color[rgb]{1,0.3,0}}%
      \expandafter\def\csname LT8\endcsname{\color[rgb]{0.5,0.5,0.5}}%
    \else
      \def\colorrgb#1{\color{black}}%
      \def\colorgray#1{\color[gray]{#1}}%
      \expandafter\def\csname LTw\endcsname{\color{white}}%
      \expandafter\def\csname LTb\endcsname{\color{black}}%
      \expandafter\def\csname LTa\endcsname{\color{black}}%
      \expandafter\def\csname LT0\endcsname{\color{black}}%
      \expandafter\def\csname LT1\endcsname{\color{black}}%
      \expandafter\def\csname LT2\endcsname{\color{black}}%
      \expandafter\def\csname LT3\endcsname{\color{black}}%
      \expandafter\def\csname LT4\endcsname{\color{black}}%
      \expandafter\def\csname LT5\endcsname{\color{black}}%
      \expandafter\def\csname LT6\endcsname{\color{black}}%
      \expandafter\def\csname LT7\endcsname{\color{black}}%
      \expandafter\def\csname LT8\endcsname{\color{black}}%
    \fi
  \fi
    \setlength{\unitlength}{0.0500bp}%
    \ifx\gptboxheight\undefined%
      \newlength{\gptboxheight}%
      \newlength{\gptboxwidth}%
      \newsavebox{\gptboxtext}%
    \fi%
    \setlength{\fboxrule}{0.5pt}%
    \setlength{\fboxsep}{1pt}%
    \definecolor{tbcol}{rgb}{1,1,1}%
\begin{picture}(7200.00,5040.00)%
    \gplgaddtomacro\gplbacktext{%
      \csname LTb\endcsname
      \put(458,518){\makebox(0,0)[r]{\strut{}$0$}}%
      \csname LTb\endcsname
      \put(458,1954){\makebox(0,0)[r]{\strut{}$1$}}%
      \csname LTb\endcsname
      \put(458,3391){\makebox(0,0)[r]{\strut{}$2$}}%
      \csname LTb\endcsname
      \put(458,4828){\makebox(0,0)[r]{\strut{}$3$}}%
      \csname LTb\endcsname
      \put(565,326){\makebox(0,0){\strut{}$0$}}%
      \csname LTb\endcsname
      \put(1764,326){\makebox(0,0){\strut{}$0.2$}}%
      \csname LTb\endcsname
      \put(2963,326){\makebox(0,0){\strut{}$0.4$}}%
      \csname LTb\endcsname
      \put(4162,326){\makebox(0,0){\strut{}$0.6$}}%
      \csname LTb\endcsname
      \put(5361,326){\makebox(0,0){\strut{}$0.8$}}%
      \csname LTb\endcsname
      \put(6559,326){\makebox(0,0){\strut{}$1$}}%
    }%
    \gplgaddtomacro\gplfronttext{%
      \csname LTb\endcsname
      \put(6035,1266){\makebox(0,0)[r]{\strut{}SPIM experiment}}%
      \csname LTb\endcsname
      \put(6035,1074){\makebox(0,0)[r]{\strut{}METIS}}%
      \csname LTb\endcsname
      \put(6035,882){\makebox(0,0)[r]{\strut{}SPIM simulation}}%
      \csname LTb\endcsname
      \put(6035,690){\makebox(0,0)[r]{\strut{}Theory}}%
      \csname LTb\endcsname
      \put(176,2673){\rotatebox{-270.00}{\makebox(0,0){\strut{}$C^*$ $(\times 1000)$}}}%
      \csname LTb\endcsname
      \put(3712,134){\makebox(0,0){\strut{}$p$}}%
    }%
    \gplgaddtomacro\gplbacktext{%
      \csname LTb\endcsname
      \put(1014,2681){\makebox(0,0)[r]{\strut{}$0$}}%
      \csname LTb\endcsname
      \put(1014,3679){\makebox(0,0)[r]{\strut{}$1$}}%
      \csname LTb\endcsname
      \put(1014,4677){\makebox(0,0)[r]{\strut{}$2$}}%
      \csname LTb\endcsname
      \put(1191,2527){\makebox(0,0){\strut{}$0$}}%
      \csname LTb\endcsname
      \put(2420,2527){\makebox(0,0){\strut{}$0.5$}}%
      \csname LTb\endcsname
      \put(3649,2527){\makebox(0,0){\strut{}$1$}}%
    }%
    \gplgaddtomacro\gplfronttext{%
      \csname LTb\endcsname
      \put(838,3679){\rotatebox{-270.00}{\makebox(0,0){\strut{}$\Delta$ $(\times 100)$}}}%
      \csname LTb\endcsname
      \put(2420,2393){\makebox(0,0){\strut{}$p$}}%
    }%
    \gplbacktext
    \put(0,0){\includegraphics[width={360.00bp},height={252.00bp}]{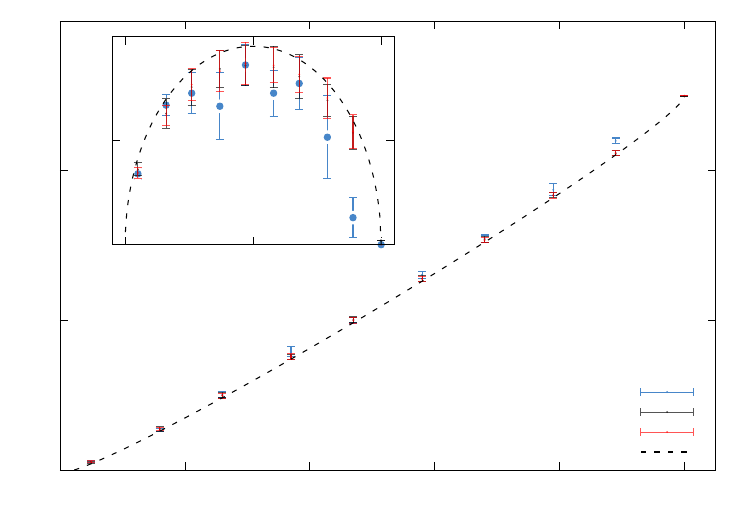}}%
    \gplfronttext
  \end{picture}%
\endgroup

%% file: main.bbl
\begin{thebibliography}{41}%
\makeatletter
\providecommand \@ifxundefined [1]{%
 \@ifx{#1\undefined}
}%
\providecommand \@ifnum [1]{%
 \ifnum #1\expandafter \@firstoftwo
 \else \expandafter \@secondoftwo
 \fi
}%
\providecommand \@ifx [1]{%
 \ifx #1\expandafter \@firstoftwo
 \else \expandafter \@secondoftwo
 \fi
}%
\providecommand \natexlab [1]{#1}%
\providecommand \enquote  [1]{``#1''}%
\providecommand \bibnamefont  [1]{#1}%
\providecommand \bibfnamefont [1]{#1}%
\providecommand \citenamefont [1]{#1}%
\providecommand \href@noop [0]{\@secondoftwo}%
\providecommand \href [0]{\begingroup \@sanitize@url \@href}%
\providecommand \@href[1]{\@@startlink{#1}\@@href}%
\providecommand \@@href[1]{\endgroup#1\@@endlink}%
\providecommand \@sanitize@url [0]{\catcode `\\12\catcode `\$12\catcode
  `\&12\catcode `\#12\catcode `\^12\catcode `\_12\catcode `\%12\relax}%
\providecommand \@@startlink[1]{}%
\providecommand \@@endlink[0]{}%
\providecommand \url  [0]{\begingroup\@sanitize@url \@url }%
\providecommand \@url [1]{\endgroup\@href {#1}{\urlprefix }}%
\providecommand \urlprefix  [0]{URL }%
\providecommand \Eprint [0]{\href }%
\providecommand \doibase [0]{https://doi.org/}%
\providecommand \selectlanguage [0]{\@gobble}%
\providecommand \bibinfo  [0]{\@secondoftwo}%
\providecommand \bibfield  [0]{\@secondoftwo}%
\providecommand \translation [1]{[#1]}%
\providecommand \BibitemOpen [0]{}%
\providecommand \bibitemStop [0]{}%
\providecommand \bibitemNoStop [0]{.\EOS\space}%
\providecommand \EOS [0]{\spacefactor3000\relax}%
\providecommand \BibitemShut  [1]{\csname bibitem#1\endcsname}%
\let\auto@bib@innerbib\@empty
\bibitem [{\citenamefont {M{\'e}zard}\ \emph {et~al.}(1987)\citenamefont
  {M{\'e}zard}, \citenamefont {Parisi},\ and\ \citenamefont
  {Virasoro}}]{mezard1987spin}%
  \BibitemOpen
  \bibfield  {author} {\bibinfo {author} {\bibfnamefont {M.}~\bibnamefont
  {M{\'e}zard}}, \bibinfo {author} {\bibfnamefont {G.}~\bibnamefont {Parisi}},\
  and\ \bibinfo {author} {\bibfnamefont {M.~A.}\ \bibnamefont {Virasoro}},\
  }\href@noop {} {\emph {\bibinfo {title} {Spin glass theory and beyond: An
  Introduction to the Replica Method and Its Applications}}},\ Vol.~\bibinfo
  {volume} {9}\ (\bibinfo  {publisher} {World Scientific Publishing Company},\
  \bibinfo {year} {1987})\BibitemShut {NoStop}%
\bibitem [{\citenamefont {M\'ezard}\ and\ \citenamefont
  {Montanari}(2009)}]{mezard2009information}%
  \BibitemOpen
  \bibfield  {author} {\bibinfo {author} {\bibfnamefont {M.}~\bibnamefont
  {M\'ezard}}\ and\ \bibinfo {author} {\bibfnamefont {A.}~\bibnamefont
  {Montanari}},\ }\href@noop {} {\emph {\bibinfo {title} {Information, physics,
  and computation}}}\ (\bibinfo  {publisher} {Oxford University Press},\
  \bibinfo {year} {2009})\BibitemShut {NoStop}%
\bibitem [{\citenamefont {Lucas}(2014)}]{lucas_ising_2014}%
  \BibitemOpen
  \bibfield  {author} {\bibinfo {author} {\bibfnamefont {A.}~\bibnamefont
  {Lucas}},\ }\bibfield  {journal} {\bibinfo  {journal} {Frontiers in Physics}\
  }\textbf {\bibinfo {volume} {2}},\ \href
  {https://doi.org/10.3389/fphy.2014.00005} {10.3389/fphy.2014.00005} (\bibinfo
  {year} {2014}),\ \bibinfo {note} {publisher: Frontiers}\BibitemShut {NoStop}%
\bibitem [{\citenamefont {Scholl}\ \emph {et~al.}(2021)\citenamefont {Scholl},
  \citenamefont {Schuler}, \citenamefont {Williams}, \citenamefont
  {Eberharter}, \citenamefont {Barredo}, \citenamefont {Schymik}, \citenamefont
  {Lienhard}, \citenamefont {Henry}, \citenamefont {Lang}, \citenamefont
  {Lahaye}, \citenamefont {Läuchli},\ and\ \citenamefont
  {Browaeys}}]{scholl_quantum_2021}%
  \BibitemOpen
  \bibfield  {author} {\bibinfo {author} {\bibfnamefont {P.}~\bibnamefont
  {Scholl}}, \bibinfo {author} {\bibfnamefont {M.}~\bibnamefont {Schuler}},
  \bibinfo {author} {\bibfnamefont {H.~J.}\ \bibnamefont {Williams}}, \bibinfo
  {author} {\bibfnamefont {A.~A.}\ \bibnamefont {Eberharter}}, \bibinfo
  {author} {\bibfnamefont {D.}~\bibnamefont {Barredo}}, \bibinfo {author}
  {\bibfnamefont {K.-N.}\ \bibnamefont {Schymik}}, \bibinfo {author}
  {\bibfnamefont {V.}~\bibnamefont {Lienhard}}, \bibinfo {author}
  {\bibfnamefont {L.-P.}\ \bibnamefont {Henry}}, \bibinfo {author}
  {\bibfnamefont {T.~C.}\ \bibnamefont {Lang}}, \bibinfo {author}
  {\bibfnamefont {T.}~\bibnamefont {Lahaye}}, \bibinfo {author} {\bibfnamefont
  {A.~M.}\ \bibnamefont {Läuchli}},\ and\ \bibinfo {author} {\bibfnamefont
  {A.}~\bibnamefont {Browaeys}},\ }\href
  {https://doi.org/10.1038/s41586-021-03585-1} {\bibfield  {journal} {\bibinfo
  {journal} {Nature}\ }\textbf {\bibinfo {volume} {595}},\ \bibinfo {pages}
  {233} (\bibinfo {year} {2021})},\ \bibinfo {note} {publisher: Nature
  Publishing Group}\BibitemShut {NoStop}%
\bibitem [{\citenamefont {Kim}\ \emph {et~al.}(2010)\citenamefont {Kim},
  \citenamefont {Chang}, \citenamefont {Korenblit}, \citenamefont {Islam},
  \citenamefont {Edwards}, \citenamefont {Freericks}, \citenamefont {Lin},
  \citenamefont {Duan},\ and\ \citenamefont {Monroe}}]{kim_quantum_2010}%
  \BibitemOpen
  \bibfield  {author} {\bibinfo {author} {\bibfnamefont {K.}~\bibnamefont
  {Kim}}, \bibinfo {author} {\bibfnamefont {M.-S.}\ \bibnamefont {Chang}},
  \bibinfo {author} {\bibfnamefont {S.}~\bibnamefont {Korenblit}}, \bibinfo
  {author} {\bibfnamefont {R.}~\bibnamefont {Islam}}, \bibinfo {author}
  {\bibfnamefont {E.~E.}\ \bibnamefont {Edwards}}, \bibinfo {author}
  {\bibfnamefont {J.~K.}\ \bibnamefont {Freericks}}, \bibinfo {author}
  {\bibfnamefont {G.-D.}\ \bibnamefont {Lin}}, \bibinfo {author} {\bibfnamefont
  {L.-M.}\ \bibnamefont {Duan}},\ and\ \bibinfo {author} {\bibfnamefont
  {C.}~\bibnamefont {Monroe}},\ }\href {https://doi.org/10.1038/nature09071}
  {\bibfield  {journal} {\bibinfo  {journal} {Nature}\ }\textbf {\bibinfo
  {volume} {465}},\ \bibinfo {pages} {590} (\bibinfo {year} {2010})},\ \bibinfo
  {note} {publisher: Nature Publishing Group}\BibitemShut {NoStop}%
\bibitem [{\citenamefont {Ma}\ \emph {et~al.}(2011)\citenamefont {Ma},
  \citenamefont {Dakic}, \citenamefont {Naylor}, \citenamefont {Zeilinger},\
  and\ \citenamefont {Walther}}]{ma_quantum_2011}%
  \BibitemOpen
  \bibfield  {author} {\bibinfo {author} {\bibfnamefont {X.-s.}\ \bibnamefont
  {Ma}}, \bibinfo {author} {\bibfnamefont {B.}~\bibnamefont {Dakic}}, \bibinfo
  {author} {\bibfnamefont {W.}~\bibnamefont {Naylor}}, \bibinfo {author}
  {\bibfnamefont {A.}~\bibnamefont {Zeilinger}},\ and\ \bibinfo {author}
  {\bibfnamefont {P.}~\bibnamefont {Walther}},\ }\href
  {https://doi.org/10.1038/nphys1919} {\bibfield  {journal} {\bibinfo
  {journal} {Nature Physics}\ }\textbf {\bibinfo {volume} {7}},\ \bibinfo
  {pages} {399} (\bibinfo {year} {2011})},\ \bibinfo {note} {publisher: Nature
  Publishing Group}\BibitemShut {NoStop}%
\bibitem [{\citenamefont {Johnson}\ \emph {et~al.}(2011)\citenamefont
  {Johnson}, \citenamefont {Amin}, \citenamefont {Gildert}, \citenamefont
  {Lanting}, \citenamefont {Hamze}, \citenamefont {Dickson}, \citenamefont
  {Harris}, \citenamefont {Berkley}, \citenamefont {Johansson}, \citenamefont
  {Bunyk}, \citenamefont {Chapple}, \citenamefont {Enderud}, \citenamefont
  {Hilton}, \citenamefont {Karimi}, \citenamefont {Ladizinsky}, \citenamefont
  {Ladizinsky}, \citenamefont {Oh}, \citenamefont {Perminov}, \citenamefont
  {Rich}, \citenamefont {Thom}, \citenamefont {Tolkacheva}, \citenamefont
  {Truncik}, \citenamefont {Uchaikin}, \citenamefont {Wang}, \citenamefont
  {Wilson},\ and\ \citenamefont {Rose}}]{johnson_quantum_2011}%
  \BibitemOpen
  \bibfield  {author} {\bibinfo {author} {\bibfnamefont {M.~W.}\ \bibnamefont
  {Johnson}}, \bibinfo {author} {\bibfnamefont {M.~H.~S.}\ \bibnamefont
  {Amin}}, \bibinfo {author} {\bibfnamefont {S.}~\bibnamefont {Gildert}},
  \bibinfo {author} {\bibfnamefont {T.}~\bibnamefont {Lanting}}, \bibinfo
  {author} {\bibfnamefont {F.}~\bibnamefont {Hamze}}, \bibinfo {author}
  {\bibfnamefont {N.}~\bibnamefont {Dickson}}, \bibinfo {author} {\bibfnamefont
  {R.}~\bibnamefont {Harris}}, \bibinfo {author} {\bibfnamefont {A.~J.}\
  \bibnamefont {Berkley}}, \bibinfo {author} {\bibfnamefont {J.}~\bibnamefont
  {Johansson}}, \bibinfo {author} {\bibfnamefont {P.}~\bibnamefont {Bunyk}},
  \bibinfo {author} {\bibfnamefont {E.~M.}\ \bibnamefont {Chapple}}, \bibinfo
  {author} {\bibfnamefont {C.}~\bibnamefont {Enderud}}, \bibinfo {author}
  {\bibfnamefont {J.~P.}\ \bibnamefont {Hilton}}, \bibinfo {author}
  {\bibfnamefont {K.}~\bibnamefont {Karimi}}, \bibinfo {author} {\bibfnamefont
  {E.}~\bibnamefont {Ladizinsky}}, \bibinfo {author} {\bibfnamefont
  {N.}~\bibnamefont {Ladizinsky}}, \bibinfo {author} {\bibfnamefont
  {T.}~\bibnamefont {Oh}}, \bibinfo {author} {\bibfnamefont {I.}~\bibnamefont
  {Perminov}}, \bibinfo {author} {\bibfnamefont {C.}~\bibnamefont {Rich}},
  \bibinfo {author} {\bibfnamefont {M.~C.}\ \bibnamefont {Thom}}, \bibinfo
  {author} {\bibfnamefont {E.}~\bibnamefont {Tolkacheva}}, \bibinfo {author}
  {\bibfnamefont {C.~J.~S.}\ \bibnamefont {Truncik}}, \bibinfo {author}
  {\bibfnamefont {S.}~\bibnamefont {Uchaikin}}, \bibinfo {author}
  {\bibfnamefont {J.}~\bibnamefont {Wang}}, \bibinfo {author} {\bibfnamefont
  {B.}~\bibnamefont {Wilson}},\ and\ \bibinfo {author} {\bibfnamefont
  {G.}~\bibnamefont {Rose}},\ }\href {https://doi.org/10.1038/nature10012}
  {\bibfield  {journal} {\bibinfo  {journal} {Nature}\ }\textbf {\bibinfo
  {volume} {473}},\ \bibinfo {pages} {194} (\bibinfo {year} {2011})},\ \bibinfo
  {note} {publisher: Nature Publishing Group}\BibitemShut {NoStop}%
\bibitem [{\citenamefont {Berloff}\ \emph {et~al.}(2017)\citenamefont
  {Berloff}, \citenamefont {Silva}, \citenamefont {Kalinin}, \citenamefont
  {Askitopoulos}, \citenamefont {Töpfer}, \citenamefont {Cilibrizzi},
  \citenamefont {Langbein},\ and\ \citenamefont
  {Lagoudakis}}]{berloff_realizing_2017}%
  \BibitemOpen
  \bibfield  {author} {\bibinfo {author} {\bibfnamefont {N.~G.}\ \bibnamefont
  {Berloff}}, \bibinfo {author} {\bibfnamefont {M.}~\bibnamefont {Silva}},
  \bibinfo {author} {\bibfnamefont {K.}~\bibnamefont {Kalinin}}, \bibinfo
  {author} {\bibfnamefont {A.}~\bibnamefont {Askitopoulos}}, \bibinfo {author}
  {\bibfnamefont {J.~D.}\ \bibnamefont {Töpfer}}, \bibinfo {author}
  {\bibfnamefont {P.}~\bibnamefont {Cilibrizzi}}, \bibinfo {author}
  {\bibfnamefont {W.}~\bibnamefont {Langbein}},\ and\ \bibinfo {author}
  {\bibfnamefont {P.~G.}\ \bibnamefont {Lagoudakis}},\ }\href
  {https://doi.org/10.1038/nmat4971} {\bibfield  {journal} {\bibinfo  {journal}
  {Nature Materials}\ }\textbf {\bibinfo {volume} {16}},\ \bibinfo {pages}
  {1120} (\bibinfo {year} {2017})},\ \bibinfo {note} {publisher: Nature
  Publishing Group}\BibitemShut {NoStop}%
\bibitem [{\citenamefont {Ohadi}\ \emph {et~al.}(2017)\citenamefont {Ohadi},
  \citenamefont {Ramsay}, \citenamefont {Sigurdsson}, \citenamefont {del
  Valle-Inclan~Redondo}, \citenamefont {Tsintzos}, \citenamefont {Hatzopoulos},
  \citenamefont {Liew}, \citenamefont {Shelykh}, \citenamefont {Rubo},
  \citenamefont {Savvidis},\ and\ \citenamefont {Baumberg}}]{ohadi_spin_2017}%
  \BibitemOpen
  \bibfield  {author} {\bibinfo {author} {\bibfnamefont {H.}~\bibnamefont
  {Ohadi}}, \bibinfo {author} {\bibfnamefont {A.}~\bibnamefont {Ramsay}},
  \bibinfo {author} {\bibfnamefont {H.}~\bibnamefont {Sigurdsson}}, \bibinfo
  {author} {\bibfnamefont {Y.}~\bibnamefont {del Valle-Inclan~Redondo}},
  \bibinfo {author} {\bibfnamefont {S.}~\bibnamefont {Tsintzos}}, \bibinfo
  {author} {\bibfnamefont {Z.}~\bibnamefont {Hatzopoulos}}, \bibinfo {author}
  {\bibfnamefont {T.}~\bibnamefont {Liew}}, \bibinfo {author} {\bibfnamefont
  {I.}~\bibnamefont {Shelykh}}, \bibinfo {author} {\bibfnamefont
  {Y.}~\bibnamefont {Rubo}}, \bibinfo {author} {\bibfnamefont {P.}~\bibnamefont
  {Savvidis}},\ and\ \bibinfo {author} {\bibfnamefont {J.}~\bibnamefont
  {Baumberg}},\ }\href {https://doi.org/10.1103/PhysRevLett.119.067401}
  {\bibfield  {journal} {\bibinfo  {journal} {Physical Review Letters}\
  }\textbf {\bibinfo {volume} {119}},\ \bibinfo {pages} {067401} (\bibinfo
  {year} {2017})},\ \bibinfo {note} {publisher: American Physical
  Society}\BibitemShut {NoStop}%
\bibitem [{\citenamefont {Kalinin}\ \emph {et~al.}(2020)\citenamefont
  {Kalinin}, \citenamefont {Amo}, \citenamefont {Bloch},\ and\ \citenamefont
  {Berloff}}]{kalinin_polaritonic_2020}%
  \BibitemOpen
  \bibfield  {author} {\bibinfo {author} {\bibfnamefont {K.~P.}\ \bibnamefont
  {Kalinin}}, \bibinfo {author} {\bibfnamefont {A.}~\bibnamefont {Amo}},
  \bibinfo {author} {\bibfnamefont {J.}~\bibnamefont {Bloch}},\ and\ \bibinfo
  {author} {\bibfnamefont {N.~G.}\ \bibnamefont {Berloff}},\ }\href
  {https://doi.org/10.1515/nanoph-2020-0162} {\bibfield  {journal} {\bibinfo
  {journal} {Nanophotonics}\ }\textbf {\bibinfo {volume} {9}},\ \bibinfo
  {pages} {4127} (\bibinfo {year} {2020})},\ \bibinfo {note} {publisher: De
  Gruyter}\BibitemShut {NoStop}%
\bibitem [{\citenamefont {Alyatkin}\ \emph {et~al.}(2020)\citenamefont
  {Alyatkin}, \citenamefont {Töpfer}, \citenamefont {Askitopoulos},
  \citenamefont {Sigurdsson},\ and\ \citenamefont
  {Lagoudakis}}]{alyatkin_optical_2020}%
  \BibitemOpen
  \bibfield  {author} {\bibinfo {author} {\bibfnamefont {S.}~\bibnamefont
  {Alyatkin}}, \bibinfo {author} {\bibfnamefont {J.}~\bibnamefont {Töpfer}},
  \bibinfo {author} {\bibfnamefont {A.}~\bibnamefont {Askitopoulos}}, \bibinfo
  {author} {\bibfnamefont {H.}~\bibnamefont {Sigurdsson}},\ and\ \bibinfo
  {author} {\bibfnamefont {P.}~\bibnamefont {Lagoudakis}},\ }\href
  {https://doi.org/10.1103/PhysRevLett.124.207402} {\bibfield  {journal}
  {\bibinfo  {journal} {Physical Review Letters}\ }\textbf {\bibinfo {volume}
  {124}},\ \bibinfo {pages} {207402} (\bibinfo {year} {2020})},\ \bibinfo
  {note} {publisher: American Physical Society}\BibitemShut {NoStop}%
\bibitem [{\citenamefont {Borders}\ \emph {et~al.}(2019)\citenamefont
  {Borders}, \citenamefont {Pervaiz}, \citenamefont {Fukami}, \citenamefont
  {Camsari}, \citenamefont {Ohno},\ and\ \citenamefont
  {Datta}}]{borders_integer_2019}%
  \BibitemOpen
  \bibfield  {author} {\bibinfo {author} {\bibfnamefont {W.~A.}\ \bibnamefont
  {Borders}}, \bibinfo {author} {\bibfnamefont {A.~Z.}\ \bibnamefont
  {Pervaiz}}, \bibinfo {author} {\bibfnamefont {S.}~\bibnamefont {Fukami}},
  \bibinfo {author} {\bibfnamefont {K.~Y.}\ \bibnamefont {Camsari}}, \bibinfo
  {author} {\bibfnamefont {H.}~\bibnamefont {Ohno}},\ and\ \bibinfo {author}
  {\bibfnamefont {S.}~\bibnamefont {Datta}},\ }\href
  {https://doi.org/10.1038/s41586-019-1557-9} {\bibfield  {journal} {\bibinfo
  {journal} {Nature}\ }\textbf {\bibinfo {volume} {573}},\ \bibinfo {pages}
  {390} (\bibinfo {year} {2019})},\ \bibinfo {note} {publisher: Nature
  Publishing Group}\BibitemShut {NoStop}%
\bibitem [{\citenamefont {Cai}\ \emph {et~al.}(2020)\citenamefont {Cai},
  \citenamefont {Kumar}, \citenamefont {Van~Vaerenbergh}, \citenamefont
  {Sheng}, \citenamefont {Liu}, \citenamefont {Li}, \citenamefont {Liu},
  \citenamefont {Foltin}, \citenamefont {Yu}, \citenamefont {Xia},
  \citenamefont {Yang}, \citenamefont {Beausoleil}, \citenamefont {Lu},\ and\
  \citenamefont {Strachan}}]{cai_power-efficient_2020}%
  \BibitemOpen
  \bibfield  {author} {\bibinfo {author} {\bibfnamefont {F.}~\bibnamefont
  {Cai}}, \bibinfo {author} {\bibfnamefont {S.}~\bibnamefont {Kumar}}, \bibinfo
  {author} {\bibfnamefont {T.}~\bibnamefont {Van~Vaerenbergh}}, \bibinfo
  {author} {\bibfnamefont {X.}~\bibnamefont {Sheng}}, \bibinfo {author}
  {\bibfnamefont {R.}~\bibnamefont {Liu}}, \bibinfo {author} {\bibfnamefont
  {C.}~\bibnamefont {Li}}, \bibinfo {author} {\bibfnamefont {Z.}~\bibnamefont
  {Liu}}, \bibinfo {author} {\bibfnamefont {M.}~\bibnamefont {Foltin}},
  \bibinfo {author} {\bibfnamefont {S.}~\bibnamefont {Yu}}, \bibinfo {author}
  {\bibfnamefont {Q.}~\bibnamefont {Xia}}, \bibinfo {author} {\bibfnamefont
  {J.~J.}\ \bibnamefont {Yang}}, \bibinfo {author} {\bibfnamefont
  {R.}~\bibnamefont {Beausoleil}}, \bibinfo {author} {\bibfnamefont {W.~D.}\
  \bibnamefont {Lu}},\ and\ \bibinfo {author} {\bibfnamefont {J.~P.}\
  \bibnamefont {Strachan}},\ }\href {https://doi.org/10.1038/s41928-020-0436-6}
  {\bibfield  {journal} {\bibinfo  {journal} {Nature Electronics}\ }\textbf
  {\bibinfo {volume} {3}},\ \bibinfo {pages} {409} (\bibinfo {year} {2020})},\
  \bibinfo {note} {publisher: Nature Publishing Group}\BibitemShut {NoStop}%
\bibitem [{\citenamefont {Shukla}\ \emph {et~al.}(2014)\citenamefont {Shukla},
  \citenamefont {Parihar}, \citenamefont {Freeman}, \citenamefont {Paik},
  \citenamefont {Stone}, \citenamefont {Narayanan}, \citenamefont {Wen},
  \citenamefont {Cai}, \citenamefont {Gopalan}, \citenamefont {Engel-Herbert},
  \citenamefont {Schlom}, \citenamefont {Raychowdhury},\ and\ \citenamefont
  {Datta}}]{shukla_synchronized_2014}%
  \BibitemOpen
  \bibfield  {author} {\bibinfo {author} {\bibfnamefont {N.}~\bibnamefont
  {Shukla}}, \bibinfo {author} {\bibfnamefont {A.}~\bibnamefont {Parihar}},
  \bibinfo {author} {\bibfnamefont {E.}~\bibnamefont {Freeman}}, \bibinfo
  {author} {\bibfnamefont {H.}~\bibnamefont {Paik}}, \bibinfo {author}
  {\bibfnamefont {G.}~\bibnamefont {Stone}}, \bibinfo {author} {\bibfnamefont
  {V.}~\bibnamefont {Narayanan}}, \bibinfo {author} {\bibfnamefont
  {H.}~\bibnamefont {Wen}}, \bibinfo {author} {\bibfnamefont {Z.}~\bibnamefont
  {Cai}}, \bibinfo {author} {\bibfnamefont {V.}~\bibnamefont {Gopalan}},
  \bibinfo {author} {\bibfnamefont {R.}~\bibnamefont {Engel-Herbert}}, \bibinfo
  {author} {\bibfnamefont {D.~G.}\ \bibnamefont {Schlom}}, \bibinfo {author}
  {\bibfnamefont {A.}~\bibnamefont {Raychowdhury}},\ and\ \bibinfo {author}
  {\bibfnamefont {S.}~\bibnamefont {Datta}},\ }\href
  {https://doi.org/10.1038/srep04964} {\bibfield  {journal} {\bibinfo
  {journal} {Scientific Reports}\ }\textbf {\bibinfo {volume} {4}},\ \bibinfo
  {pages} {4964} (\bibinfo {year} {2014})},\ \bibinfo {note} {publisher: Nature
  Publishing Group}\BibitemShut {NoStop}%
\bibitem [{\citenamefont {Merolla}\ \emph {et~al.}(2014)\citenamefont
  {Merolla}, \citenamefont {Arthur}, \citenamefont {Alvarez-Icaza},
  \citenamefont {Cassidy}, \citenamefont {Sawada}, \citenamefont {Akopyan},
  \citenamefont {Jackson}, \citenamefont {Imam}, \citenamefont {Guo},
  \citenamefont {Nakamura}, \citenamefont {Brezzo}, \citenamefont {Vo},
  \citenamefont {Esser}, \citenamefont {Appuswamy}, \citenamefont {Taba},
  \citenamefont {Amir}, \citenamefont {Flickner}, \citenamefont {Risk},
  \citenamefont {Manohar},\ and\ \citenamefont {Modha}}]{merolla_million_2014}%
  \BibitemOpen
  \bibfield  {author} {\bibinfo {author} {\bibfnamefont {P.~A.}\ \bibnamefont
  {Merolla}}, \bibinfo {author} {\bibfnamefont {J.~V.}\ \bibnamefont {Arthur}},
  \bibinfo {author} {\bibfnamefont {R.}~\bibnamefont {Alvarez-Icaza}}, \bibinfo
  {author} {\bibfnamefont {A.~S.}\ \bibnamefont {Cassidy}}, \bibinfo {author}
  {\bibfnamefont {J.}~\bibnamefont {Sawada}}, \bibinfo {author} {\bibfnamefont
  {F.}~\bibnamefont {Akopyan}}, \bibinfo {author} {\bibfnamefont {B.~L.}\
  \bibnamefont {Jackson}}, \bibinfo {author} {\bibfnamefont {N.}~\bibnamefont
  {Imam}}, \bibinfo {author} {\bibfnamefont {C.}~\bibnamefont {Guo}}, \bibinfo
  {author} {\bibfnamefont {Y.}~\bibnamefont {Nakamura}}, \bibinfo {author}
  {\bibfnamefont {B.}~\bibnamefont {Brezzo}}, \bibinfo {author} {\bibfnamefont
  {I.}~\bibnamefont {Vo}}, \bibinfo {author} {\bibfnamefont {S.~K.}\
  \bibnamefont {Esser}}, \bibinfo {author} {\bibfnamefont {R.}~\bibnamefont
  {Appuswamy}}, \bibinfo {author} {\bibfnamefont {B.}~\bibnamefont {Taba}},
  \bibinfo {author} {\bibfnamefont {A.}~\bibnamefont {Amir}}, \bibinfo {author}
  {\bibfnamefont {M.~D.}\ \bibnamefont {Flickner}}, \bibinfo {author}
  {\bibfnamefont {W.~P.}\ \bibnamefont {Risk}}, \bibinfo {author}
  {\bibfnamefont {R.}~\bibnamefont {Manohar}},\ and\ \bibinfo {author}
  {\bibfnamefont {D.~S.}\ \bibnamefont {Modha}},\ }\href
  {https://doi.org/10.1126/science.1254642} {\bibfield  {journal} {\bibinfo
  {journal} {Science}\ }\textbf {\bibinfo {volume} {345}},\ \bibinfo {pages}
  {668} (\bibinfo {year} {2014})},\ \bibinfo {note} {publisher: American
  Association for the Advancement of Science}\BibitemShut {NoStop}%
\bibitem [{\citenamefont {McMahon}\ \emph {et~al.}(2016)\citenamefont
  {McMahon}, \citenamefont {Marandi}, \citenamefont {Haribara}, \citenamefont
  {Hamerly}, \citenamefont {Langrock}, \citenamefont {Tamate}, \citenamefont
  {Inagaki}, \citenamefont {Takesue}, \citenamefont {Utsunomiya}, \citenamefont
  {Aihara}, \citenamefont {Byer}, \citenamefont {Fejer}, \citenamefont
  {Mabuchi},\ and\ \citenamefont {Yamamoto}}]{mcmahon_fully_2016}%
  \BibitemOpen
  \bibfield  {author} {\bibinfo {author} {\bibfnamefont {P.~L.}\ \bibnamefont
  {McMahon}}, \bibinfo {author} {\bibfnamefont {A.}~\bibnamefont {Marandi}},
  \bibinfo {author} {\bibfnamefont {Y.}~\bibnamefont {Haribara}}, \bibinfo
  {author} {\bibfnamefont {R.}~\bibnamefont {Hamerly}}, \bibinfo {author}
  {\bibfnamefont {C.}~\bibnamefont {Langrock}}, \bibinfo {author}
  {\bibfnamefont {S.}~\bibnamefont {Tamate}}, \bibinfo {author} {\bibfnamefont
  {T.}~\bibnamefont {Inagaki}}, \bibinfo {author} {\bibfnamefont
  {H.}~\bibnamefont {Takesue}}, \bibinfo {author} {\bibfnamefont
  {S.}~\bibnamefont {Utsunomiya}}, \bibinfo {author} {\bibfnamefont
  {K.}~\bibnamefont {Aihara}}, \bibinfo {author} {\bibfnamefont {R.~L.}\
  \bibnamefont {Byer}}, \bibinfo {author} {\bibfnamefont {M.~M.}\ \bibnamefont
  {Fejer}}, \bibinfo {author} {\bibfnamefont {H.}~\bibnamefont {Mabuchi}},\
  and\ \bibinfo {author} {\bibfnamefont {Y.}~\bibnamefont {Yamamoto}},\ }\href
  {https://doi.org/10.1126/science.aah5178} {\bibfield  {journal} {\bibinfo
  {journal} {Science}\ }\textbf {\bibinfo {volume} {354}},\ \bibinfo {pages}
  {614} (\bibinfo {year} {2016})},\ \bibinfo {note} {publisher: American
  Association for the Advancement of Science}\BibitemShut {NoStop}%
\bibitem [{\citenamefont {Inagaki}\ \emph {et~al.}(2016)\citenamefont
  {Inagaki}, \citenamefont {Haribara}, \citenamefont {Igarashi}, \citenamefont
  {Sonobe}, \citenamefont {Tamate}, \citenamefont {Honjo}, \citenamefont
  {Marandi}, \citenamefont {McMahon}, \citenamefont {Umeki}, \citenamefont
  {Enbutsu}, \citenamefont {Tadanaga}, \citenamefont {Takenouchi},
  \citenamefont {Aihara}, \citenamefont {Kawarabayashi}, \citenamefont {Inoue},
  \citenamefont {Utsunomiya},\ and\ \citenamefont
  {Takesue}}]{inagaki_coherent_2016}%
  \BibitemOpen
  \bibfield  {author} {\bibinfo {author} {\bibfnamefont {T.}~\bibnamefont
  {Inagaki}}, \bibinfo {author} {\bibfnamefont {Y.}~\bibnamefont {Haribara}},
  \bibinfo {author} {\bibfnamefont {K.}~\bibnamefont {Igarashi}}, \bibinfo
  {author} {\bibfnamefont {T.}~\bibnamefont {Sonobe}}, \bibinfo {author}
  {\bibfnamefont {S.}~\bibnamefont {Tamate}}, \bibinfo {author} {\bibfnamefont
  {T.}~\bibnamefont {Honjo}}, \bibinfo {author} {\bibfnamefont
  {A.}~\bibnamefont {Marandi}}, \bibinfo {author} {\bibfnamefont {P.~L.}\
  \bibnamefont {McMahon}}, \bibinfo {author} {\bibfnamefont {T.}~\bibnamefont
  {Umeki}}, \bibinfo {author} {\bibfnamefont {K.}~\bibnamefont {Enbutsu}},
  \bibinfo {author} {\bibfnamefont {O.}~\bibnamefont {Tadanaga}}, \bibinfo
  {author} {\bibfnamefont {H.}~\bibnamefont {Takenouchi}}, \bibinfo {author}
  {\bibfnamefont {K.}~\bibnamefont {Aihara}}, \bibinfo {author} {\bibfnamefont
  {K.-i.}\ \bibnamefont {Kawarabayashi}}, \bibinfo {author} {\bibfnamefont
  {K.}~\bibnamefont {Inoue}}, \bibinfo {author} {\bibfnamefont
  {S.}~\bibnamefont {Utsunomiya}},\ and\ \bibinfo {author} {\bibfnamefont
  {H.}~\bibnamefont {Takesue}},\ }\href
  {https://doi.org/10.1126/science.aah4243} {\bibfield  {journal} {\bibinfo
  {journal} {Science}\ }\textbf {\bibinfo {volume} {354}},\ \bibinfo {pages}
  {603} (\bibinfo {year} {2016})},\ \bibinfo {note} {publisher: American
  Association for the Advancement of Science}\BibitemShut {NoStop}%
\bibitem [{\citenamefont {Pierangeli}\ \emph {et~al.}(2019)\citenamefont
  {Pierangeli}, \citenamefont {Marcucci},\ and\ \citenamefont
  {Conti}}]{pierangeli_large-scale_2019}%
  \BibitemOpen
  \bibfield  {author} {\bibinfo {author} {\bibfnamefont {D.}~\bibnamefont
  {Pierangeli}}, \bibinfo {author} {\bibfnamefont {G.}~\bibnamefont
  {Marcucci}},\ and\ \bibinfo {author} {\bibfnamefont {C.}~\bibnamefont
  {Conti}},\ }\href {https://doi.org/10.1103/PhysRevLett.122.213902} {\bibfield
   {journal} {\bibinfo  {journal} {Physical Review Letters}\ }\textbf {\bibinfo
  {volume} {122}},\ \bibinfo {pages} {213902} (\bibinfo {year} {2019})},\
  \bibinfo {note} {publisher: American Physical Society}\BibitemShut {NoStop}%
\bibitem [{\citenamefont {Honjo}\ \emph {et~al.}(2021)\citenamefont {Honjo},
  \citenamefont {Sonobe}, \citenamefont {Inaba}, \citenamefont {Inagaki},
  \citenamefont {Ikuta}, \citenamefont {Yamada}, \citenamefont {Kazama},
  \citenamefont {Enbutsu}, \citenamefont {Umeki}, \citenamefont {Kasahara},
  \citenamefont {Kawarabayashi},\ and\ \citenamefont
  {Takesue}}]{honjo_100000-spin_2021}%
  \BibitemOpen
  \bibfield  {author} {\bibinfo {author} {\bibfnamefont {T.}~\bibnamefont
  {Honjo}}, \bibinfo {author} {\bibfnamefont {T.}~\bibnamefont {Sonobe}},
  \bibinfo {author} {\bibfnamefont {K.}~\bibnamefont {Inaba}}, \bibinfo
  {author} {\bibfnamefont {T.}~\bibnamefont {Inagaki}}, \bibinfo {author}
  {\bibfnamefont {T.}~\bibnamefont {Ikuta}}, \bibinfo {author} {\bibfnamefont
  {Y.}~\bibnamefont {Yamada}}, \bibinfo {author} {\bibfnamefont
  {T.}~\bibnamefont {Kazama}}, \bibinfo {author} {\bibfnamefont
  {K.}~\bibnamefont {Enbutsu}}, \bibinfo {author} {\bibfnamefont
  {T.}~\bibnamefont {Umeki}}, \bibinfo {author} {\bibfnamefont
  {R.}~\bibnamefont {Kasahara}}, \bibinfo {author} {\bibfnamefont {K.-i.}\
  \bibnamefont {Kawarabayashi}},\ and\ \bibinfo {author} {\bibfnamefont
  {H.}~\bibnamefont {Takesue}},\ }\href
  {https://doi.org/10.1126/sciadv.abh0952} {\bibfield  {journal} {\bibinfo
  {journal} {Science Advances}\ }\textbf {\bibinfo {volume} {7}},\ \bibinfo
  {pages} {eabh0952} (\bibinfo {year} {2021})},\ \bibinfo {note} {publisher:
  American Association for the Advancement of Science}\BibitemShut {NoStop}%
\bibitem [{\citenamefont {Sun}\ \emph {et~al.}(2022)\citenamefont {Sun},
  \citenamefont {Zhang}, \citenamefont {Liu}, \citenamefont {Liu},\ and\
  \citenamefont {He}}]{sun_quadrature_2022}%
  \BibitemOpen
  \bibfield  {author} {\bibinfo {author} {\bibfnamefont {W.}~\bibnamefont
  {Sun}}, \bibinfo {author} {\bibfnamefont {W.}~\bibnamefont {Zhang}}, \bibinfo
  {author} {\bibfnamefont {Y.}~\bibnamefont {Liu}}, \bibinfo {author}
  {\bibfnamefont {Q.}~\bibnamefont {Liu}},\ and\ \bibinfo {author}
  {\bibfnamefont {Z.}~\bibnamefont {He}},\ }\href
  {https://doi.org/10.1364/OL.446789} {\bibfield  {journal} {\bibinfo
  {journal} {Optics Letters}\ }\textbf {\bibinfo {volume} {47}},\ \bibinfo
  {pages} {1498} (\bibinfo {year} {2022})},\ \bibinfo {note} {publisher: Optica
  Publishing Group}\BibitemShut {NoStop}%
\bibitem [{\citenamefont {Pierangeli}\ \emph
  {et~al.}(2020{\natexlab{a}})\citenamefont {Pierangeli}, \citenamefont
  {Marcucci},\ and\ \citenamefont {Conti}}]{pierangeli_adiabatic_2020}%
  \BibitemOpen
  \bibfield  {author} {\bibinfo {author} {\bibfnamefont {D.}~\bibnamefont
  {Pierangeli}}, \bibinfo {author} {\bibfnamefont {G.}~\bibnamefont
  {Marcucci}},\ and\ \bibinfo {author} {\bibfnamefont {C.}~\bibnamefont
  {Conti}},\ }\href {https://doi.org/10.1364/OPTICA.398000} {\bibfield
  {journal} {\bibinfo  {journal} {Optica}\ }\textbf {\bibinfo {volume} {7}},\
  \bibinfo {pages} {1535} (\bibinfo {year} {2020}{\natexlab{a}})},\ \bibinfo
  {note} {publisher: Optica Publishing Group}\BibitemShut {NoStop}%
\bibitem [{\citenamefont {Pierangeli}\ \emph
  {et~al.}(2020{\natexlab{b}})\citenamefont {Pierangeli}, \citenamefont
  {Marcucci}, \citenamefont {Brunner},\ and\ \citenamefont
  {Conti}}]{pierangeli_noise-enhanced_2020}%
  \BibitemOpen
  \bibfield  {author} {\bibinfo {author} {\bibfnamefont {D.}~\bibnamefont
  {Pierangeli}}, \bibinfo {author} {\bibfnamefont {G.}~\bibnamefont
  {Marcucci}}, \bibinfo {author} {\bibfnamefont {D.}~\bibnamefont {Brunner}},\
  and\ \bibinfo {author} {\bibfnamefont {C.}~\bibnamefont {Conti}},\ }\href
  {https://doi.org/10.1515/nanoph-2020-0119} {\bibfield  {journal} {\bibinfo
  {journal} {Nanophotonics}\ }\textbf {\bibinfo {volume} {9}},\ \bibinfo
  {pages} {4109} (\bibinfo {year} {2020}{\natexlab{b}})},\ \bibinfo {note}
  {publisher: De Gruyter}\BibitemShut {NoStop}%
\bibitem [{\citenamefont {Kumar}\ \emph {et~al.}(2020)\citenamefont {Kumar},
  \citenamefont {Zhang},\ and\ \citenamefont {Huang}}]{kumar_large-scale_2020}%
  \BibitemOpen
  \bibfield  {author} {\bibinfo {author} {\bibfnamefont {S.}~\bibnamefont
  {Kumar}}, \bibinfo {author} {\bibfnamefont {H.}~\bibnamefont {Zhang}},\ and\
  \bibinfo {author} {\bibfnamefont {Y.-P.}\ \bibnamefont {Huang}},\ }\href
  {https://doi.org/10.1038/s42005-020-0376-5} {\bibfield  {journal} {\bibinfo
  {journal} {Communications Physics}\ }\textbf {\bibinfo {volume} {3}},\
  \bibinfo {pages} {1} (\bibinfo {year} {2020})},\ \bibinfo {note} {publisher:
  Nature Publishing Group}\BibitemShut {NoStop}%
\bibitem [{\citenamefont {Kumar}\ \emph {et~al.}(2023)\citenamefont {Kumar},
  \citenamefont {Li}, \citenamefont {Bu}, \citenamefont {Qu},\ and\
  \citenamefont {Huang}}]{kumar_observation_2023}%
  \BibitemOpen
  \bibfield  {author} {\bibinfo {author} {\bibfnamefont {S.}~\bibnamefont
  {Kumar}}, \bibinfo {author} {\bibfnamefont {Z.}~\bibnamefont {Li}}, \bibinfo
  {author} {\bibfnamefont {T.}~\bibnamefont {Bu}}, \bibinfo {author}
  {\bibfnamefont {C.}~\bibnamefont {Qu}},\ and\ \bibinfo {author}
  {\bibfnamefont {Y.}~\bibnamefont {Huang}},\ }\href
  {https://doi.org/10.1038/s42005-023-01148-6} {\bibfield  {journal} {\bibinfo
  {journal} {Communications Physics}\ }\textbf {\bibinfo {volume} {6}},\
  \bibinfo {pages} {1} (\bibinfo {year} {2023})},\ \bibinfo {note} {publisher:
  Nature Publishing Group}\BibitemShut {NoStop}%
\bibitem [{\citenamefont {Fang}\ \emph {et~al.}(2021)\citenamefont {Fang},
  \citenamefont {Huang},\ and\ \citenamefont {Ruan}}]{fang_experimental_2021}%
  \BibitemOpen
  \bibfield  {author} {\bibinfo {author} {\bibfnamefont {Y.}~\bibnamefont
  {Fang}}, \bibinfo {author} {\bibfnamefont {J.}~\bibnamefont {Huang}},\ and\
  \bibinfo {author} {\bibfnamefont {Z.}~\bibnamefont {Ruan}},\ }\href
  {https://doi.org/10.1103/PhysRevLett.127.043902} {\bibfield  {journal}
  {\bibinfo  {journal} {Physical Review Letters}\ }\textbf {\bibinfo {volume}
  {127}},\ \bibinfo {pages} {043902} (\bibinfo {year} {2021})},\ \bibinfo
  {note} {publisher: American Physical Society}\BibitemShut {NoStop}%
\bibitem [{\citenamefont {Leonetti}\ \emph {et~al.}(2021)\citenamefont
  {Leonetti}, \citenamefont {Hörmann}, \citenamefont {Leuzzi}, \citenamefont
  {Parisi},\ and\ \citenamefont {Ruocco}}]{leonetti_optical_2021}%
  \BibitemOpen
  \bibfield  {author} {\bibinfo {author} {\bibfnamefont {M.}~\bibnamefont
  {Leonetti}}, \bibinfo {author} {\bibfnamefont {E.}~\bibnamefont {Hörmann}},
  \bibinfo {author} {\bibfnamefont {L.}~\bibnamefont {Leuzzi}}, \bibinfo
  {author} {\bibfnamefont {G.}~\bibnamefont {Parisi}},\ and\ \bibinfo {author}
  {\bibfnamefont {G.}~\bibnamefont {Ruocco}},\ }\href
  {https://doi.org/10.1073/pnas.2015207118} {\bibfield  {journal} {\bibinfo
  {journal} {Proceedings of the National Academy of Sciences}\ }\textbf
  {\bibinfo {volume} {118}},\ \bibinfo {pages} {e2015207118} (\bibinfo {year}
  {2021})},\ \bibinfo {note} {publisher: Proceedings of the National Academy of
  Sciences}\BibitemShut {NoStop}%
\bibitem [{\citenamefont {Huang}\ \emph {et~al.}(2021)\citenamefont {Huang},
  \citenamefont {Fang},\ and\ \citenamefont
  {Ruan}}]{huang_antiferromagnetic_2021}%
  \BibitemOpen
  \bibfield  {author} {\bibinfo {author} {\bibfnamefont {J.}~\bibnamefont
  {Huang}}, \bibinfo {author} {\bibfnamefont {Y.}~\bibnamefont {Fang}},\ and\
  \bibinfo {author} {\bibfnamefont {Z.}~\bibnamefont {Ruan}},\ }\href
  {https://doi.org/10.1038/s42005-021-00741-x} {\bibfield  {journal} {\bibinfo
  {journal} {Communications Physics}\ }\textbf {\bibinfo {volume} {4}},\
  \bibinfo {pages} {1} (\bibinfo {year} {2021})},\ \bibinfo {note} {publisher:
  Nature Publishing Group}\BibitemShut {NoStop}%
\bibitem [{\citenamefont {Prabhakar}\ \emph {et~al.}(2022)\citenamefont
  {Prabhakar}, \citenamefont {Shah}, \citenamefont {Gautham}, \citenamefont
  {Natarajan}, \citenamefont {Ramesh}, \citenamefont {Chandrachoodan},\ and\
  \citenamefont {Tayur}}]{prabhakar_optimization_2022}%
  \BibitemOpen
  \bibfield  {author} {\bibinfo {author} {\bibfnamefont {A.}~\bibnamefont
  {Prabhakar}}, \bibinfo {author} {\bibfnamefont {P.}~\bibnamefont {Shah}},
  \bibinfo {author} {\bibfnamefont {U.}~\bibnamefont {Gautham}}, \bibinfo
  {author} {\bibfnamefont {V.}~\bibnamefont {Natarajan}}, \bibinfo {author}
  {\bibfnamefont {V.}~\bibnamefont {Ramesh}}, \bibinfo {author} {\bibfnamefont
  {N.}~\bibnamefont {Chandrachoodan}},\ and\ \bibinfo {author} {\bibfnamefont
  {S.}~\bibnamefont {Tayur}},\ }\href {https://doi.org/10.1098/rsta.2021.0409}
  {\bibfield  {journal} {\bibinfo  {journal} {Philosophical Transactions of the
  Royal Society A: Mathematical, Physical and Engineering Sciences}\ }\textbf
  {\bibinfo {volume} {381}},\ \bibinfo {pages} {20210409} (\bibinfo {year}
  {2022})},\ \bibinfo {note} {publisher: Royal Society}\BibitemShut {NoStop}%
\bibitem [{\citenamefont {Ye}\ \emph {et~al.}(2023)\citenamefont {Ye},
  \citenamefont {Zhang}, \citenamefont {Wang}, \citenamefont {Yang},
  \citenamefont {Du},\ and\ \citenamefont {He}}]{ye_photonic_2023}%
  \BibitemOpen
  \bibfield  {author} {\bibinfo {author} {\bibfnamefont {X.}~\bibnamefont
  {Ye}}, \bibinfo {author} {\bibfnamefont {W.}~\bibnamefont {Zhang}}, \bibinfo
  {author} {\bibfnamefont {S.}~\bibnamefont {Wang}}, \bibinfo {author}
  {\bibfnamefont {X.}~\bibnamefont {Yang}}, \bibinfo {author} {\bibfnamefont
  {J.}~\bibnamefont {Du}},\ and\ \bibinfo {author} {\bibfnamefont
  {Z.}~\bibnamefont {He}},\ }in\ \href
  {https://doi.org/10.1364/CLEO_AT.2023.JTh2A.32} {\emph {\bibinfo {booktitle}
  {{CLEO} 2023 (2023), paper {JTh2A}.32}}}\ (\bibinfo  {publisher} {Optica
  Publishing Group},\ \bibinfo {year} {2023})\ p.\ \bibinfo {pages}
  {JTh2A.32}\BibitemShut {NoStop}%
\bibitem [{\citenamefont {Yamashita}\ \emph {et~al.}(2023)\citenamefont
  {Yamashita}, \citenamefont {Okubo}, \citenamefont {Shimomura}, \citenamefont
  {Ogura}, \citenamefont {Tanida},\ and\ \citenamefont
  {Suzuki}}]{yamashita_low-rank_2023}%
  \BibitemOpen
  \bibfield  {author} {\bibinfo {author} {\bibfnamefont {H.}~\bibnamefont
  {Yamashita}}, \bibinfo {author} {\bibfnamefont {K.-i.}\ \bibnamefont
  {Okubo}}, \bibinfo {author} {\bibfnamefont {S.}~\bibnamefont {Shimomura}},
  \bibinfo {author} {\bibfnamefont {Y.}~\bibnamefont {Ogura}}, \bibinfo
  {author} {\bibfnamefont {J.}~\bibnamefont {Tanida}},\ and\ \bibinfo {author}
  {\bibfnamefont {H.}~\bibnamefont {Suzuki}},\ }\href
  {https://doi.org/10.1103/PhysRevLett.131.063801} {\bibfield  {journal}
  {\bibinfo  {journal} {Physical Review Letters}\ }\textbf {\bibinfo {volume}
  {131}},\ \bibinfo {pages} {063801} (\bibinfo {year} {2023})},\ \bibinfo
  {note} {publisher: American Physical Society}\BibitemShut {NoStop}%
\bibitem [{\citenamefont {Wang}\ \emph
  {et~al.}(2024{\natexlab{a}})\citenamefont {Wang}, \citenamefont {Cummins},
  \citenamefont {Syed}, \citenamefont {Stroev}, \citenamefont {Pastras},
  \citenamefont {Sakellariou}, \citenamefont {Tsintzos}, \citenamefont
  {Askitopoulos}, \citenamefont {Veraldi}, \citenamefont {Strinati},
  \citenamefont {Gentilini}, \citenamefont {Pierangeli}, \citenamefont
  {Conti},\ and\ \citenamefont {Berloff}}]{wang_efficient_2024}%
  \BibitemOpen
  \bibfield  {author} {\bibinfo {author} {\bibfnamefont {R.~Z.}\ \bibnamefont
  {Wang}}, \bibinfo {author} {\bibfnamefont {J.~S.}\ \bibnamefont {Cummins}},
  \bibinfo {author} {\bibfnamefont {M.}~\bibnamefont {Syed}}, \bibinfo {author}
  {\bibfnamefont {N.}~\bibnamefont {Stroev}}, \bibinfo {author} {\bibfnamefont
  {G.}~\bibnamefont {Pastras}}, \bibinfo {author} {\bibfnamefont
  {J.}~\bibnamefont {Sakellariou}}, \bibinfo {author} {\bibfnamefont
  {S.}~\bibnamefont {Tsintzos}}, \bibinfo {author} {\bibfnamefont
  {A.}~\bibnamefont {Askitopoulos}}, \bibinfo {author} {\bibfnamefont
  {D.}~\bibnamefont {Veraldi}}, \bibinfo {author} {\bibfnamefont {M.~C.}\
  \bibnamefont {Strinati}}, \bibinfo {author} {\bibfnamefont {S.}~\bibnamefont
  {Gentilini}}, \bibinfo {author} {\bibfnamefont {D.}~\bibnamefont
  {Pierangeli}}, \bibinfo {author} {\bibfnamefont {C.}~\bibnamefont {Conti}},\
  and\ \bibinfo {author} {\bibfnamefont {N.~G.}\ \bibnamefont {Berloff}},\
  }\href {https://doi.org/10.48550/arXiv.2406.01400} {\bibinfo {title}
  {Efficient {Computation} {Using} {Spatial}-{Photonic} {Ising} {Machines}:
  {Utilizing} {Low}-{Rank} and {Circulant} {Matrix} {Constraints}}} (\bibinfo
  {year} {2024}{\natexlab{a}}),\ \bibinfo {note} {arXiv:2406.01400 [cond-mat,
  physics:physics]}\BibitemShut {NoStop}%
\bibitem [{\citenamefont {Mattis}(1976)}]{mattis_solvable_1976}%
  \BibitemOpen
  \bibfield  {author} {\bibinfo {author} {\bibfnamefont {D.~C.}\ \bibnamefont
  {Mattis}},\ }\href {https://doi.org/10.1016/0375-9601(76)90396-0} {\bibfield
  {journal} {\bibinfo  {journal} {Physics Letters A}\ }\textbf {\bibinfo
  {volume} {56}},\ \bibinfo {pages} {421} (\bibinfo {year} {1976})}\BibitemShut
  {NoStop}%
\bibitem [{\citenamefont {Ouyang}\ \emph {et~al.}(2024)\citenamefont {Ouyang},
  \citenamefont {Liao}, \citenamefont {Ma}, \citenamefont {Kong}, \citenamefont
  {Feng}, \citenamefont {Zhang}, \citenamefont {Dong}, \citenamefont {Cui},
  \citenamefont {Liu}, \citenamefont {Zhang},\ and\ \citenamefont
  {Huang}}]{ouyang_-demand_2024}%
  \BibitemOpen
  \bibfield  {author} {\bibinfo {author} {\bibfnamefont {J.}~\bibnamefont
  {Ouyang}}, \bibinfo {author} {\bibfnamefont {Y.}~\bibnamefont {Liao}},
  \bibinfo {author} {\bibfnamefont {Z.}~\bibnamefont {Ma}}, \bibinfo {author}
  {\bibfnamefont {D.}~\bibnamefont {Kong}}, \bibinfo {author} {\bibfnamefont
  {X.}~\bibnamefont {Feng}}, \bibinfo {author} {\bibfnamefont {X.}~\bibnamefont
  {Zhang}}, \bibinfo {author} {\bibfnamefont {X.}~\bibnamefont {Dong}},
  \bibinfo {author} {\bibfnamefont {K.}~\bibnamefont {Cui}}, \bibinfo {author}
  {\bibfnamefont {F.}~\bibnamefont {Liu}}, \bibinfo {author} {\bibfnamefont
  {W.}~\bibnamefont {Zhang}},\ and\ \bibinfo {author} {\bibfnamefont
  {Y.}~\bibnamefont {Huang}},\ }\href
  {https://doi.org/10.1038/s42005-024-01658-x} {\bibfield  {journal} {\bibinfo
  {journal} {Communications Physics}\ }\textbf {\bibinfo {volume} {7}},\
  \bibinfo {pages} {1} (\bibinfo {year} {2024})},\ \bibinfo {note} {publisher:
  Nature Publishing Group}\BibitemShut {NoStop}%
\bibitem [{\citenamefont {Wang}\ \emph
  {et~al.}(2024{\natexlab{b}})\citenamefont {Wang}, \citenamefont {Zhang},
  \citenamefont {Ye},\ and\ \citenamefont {He}}]{wang_general_2024}%
  \BibitemOpen
  \bibfield  {author} {\bibinfo {author} {\bibfnamefont {S.}~\bibnamefont
  {Wang}}, \bibinfo {author} {\bibfnamefont {W.}~\bibnamefont {Zhang}},
  \bibinfo {author} {\bibfnamefont {X.}~\bibnamefont {Ye}},\ and\ \bibinfo
  {author} {\bibfnamefont {Z.}~\bibnamefont {He}},\ }\href
  {https://doi.org/10.1364/AO.521061} {\bibfield  {journal} {\bibinfo
  {journal} {Applied Optics}\ }\textbf {\bibinfo {volume} {63}},\ \bibinfo
  {pages} {2973} (\bibinfo {year} {2024}{\natexlab{b}})},\ \bibinfo {note}
  {publisher: Optica Publishing Group}\BibitemShut {NoStop}%
\bibitem [{\citenamefont {Luo}\ \emph {et~al.}(2023)\citenamefont {Luo},
  \citenamefont {Mi}, \citenamefont {Huang},\ and\ \citenamefont
  {Ruan}}]{luo_wavelength-division_2023}%
  \BibitemOpen
  \bibfield  {author} {\bibinfo {author} {\bibfnamefont {L.}~\bibnamefont
  {Luo}}, \bibinfo {author} {\bibfnamefont {Z.}~\bibnamefont {Mi}}, \bibinfo
  {author} {\bibfnamefont {J.}~\bibnamefont {Huang}},\ and\ \bibinfo {author}
  {\bibfnamefont {Z.}~\bibnamefont {Ruan}},\ }\href
  {https://doi.org/10.1126/sciadv.adg6238} {\bibfield  {journal} {\bibinfo
  {journal} {Science Advances}\ }\textbf {\bibinfo {volume} {9}},\ \bibinfo
  {pages} {eadg6238} (\bibinfo {year} {2023})},\ \bibinfo {note} {publisher:
  American Association for the Advancement of Science}\BibitemShut {NoStop}%
\bibitem [{sup()}]{supp}%
  \BibitemOpen
  \href@noop {} {}\bibinfo {note} {See Supplemental Material at [URL will be
  inserted by publisher].
  }\BibitemShut {Stop}%
\bibitem [{kar(2024)}]{karypislabmetis_2024}%
  \BibitemOpen
  \href {https://github.com/KarypisLab/METIS} {\bibinfo {title}
  {{KarypisLab}/{METIS}}} (\bibinfo {year} {2024}),\ \bibinfo {note}
  {original-date: 2020-03-02T03:37:17Z}\BibitemShut {NoStop}%
\bibitem [{\citenamefont {LaSalle}\ and\ \citenamefont
  {Karypis}(2016)}]{lasalle_parallel_2016}%
  \BibitemOpen
  \bibfield  {author} {\bibinfo {author} {\bibfnamefont {D.}~\bibnamefont
  {LaSalle}}\ and\ \bibinfo {author} {\bibfnamefont {G.}~\bibnamefont
  {Karypis}},\ }in\ \href {https://doi.org/10.1109/ICPP.2016.34} {\emph
  {\bibinfo {booktitle} {2016 45th {International} {Conference} on {Parallel}
  {Processing} ({ICPP})}}}\ (\bibinfo {year} {2016})\ pp.\ \bibinfo {pages}
  {236--241},\ \bibinfo {note} {iSSN: 2332-5690}\BibitemShut {NoStop}%
\bibitem [{\citenamefont {Fu}\ and\ \citenamefont
  {Anderson}(1986)}]{fu1986application}%
  \BibitemOpen
  \bibfield  {author} {\bibinfo {author} {\bibfnamefont {Y.}~\bibnamefont
  {Fu}}\ and\ \bibinfo {author} {\bibfnamefont {P.~W.}\ \bibnamefont
  {Anderson}},\ }\href@noop {} {\bibfield  {journal} {\bibinfo  {journal}
  {Journal of Physics A: Mathematical and General}\ }\textbf {\bibinfo {volume}
  {19}},\ \bibinfo {pages} {1605} (\bibinfo {year} {1986})}\BibitemShut
  {NoStop}%
\bibitem [{\citenamefont {Bashar}\ and\ \citenamefont
  {Shukla}(2023)}]{bashar_designing_2023}%
  \BibitemOpen
  \bibfield  {author} {\bibinfo {author} {\bibfnamefont {M.~K.}\ \bibnamefont
  {Bashar}}\ and\ \bibinfo {author} {\bibfnamefont {N.}~\bibnamefont
  {Shukla}},\ }\href {https://doi.org/10.1038/s41598-023-36531-4} {\bibfield
  {journal} {\bibinfo  {journal} {Scientific Reports}\ }\textbf {\bibinfo
  {volume} {13}},\ \bibinfo {pages} {9558} (\bibinfo {year} {2023})},\ \bibinfo
  {note} {publisher: Nature Publishing Group}\BibitemShut {NoStop}%
\bibitem [{\citenamefont {Park}\ \emph {et~al.}(2021)\citenamefont {Park},
  \citenamefont {Jeong}, \citenamefont {Kim}, \citenamefont {Lee},
  \citenamefont {Kim}, \citenamefont {Shin}, \citenamefont {Lee}, \citenamefont
  {Otsuka}, \citenamefont {Kyoung}, \citenamefont {Kim}, \citenamefont {Yang},
  \citenamefont {Park}, \citenamefont {Lee}, \citenamefont {Hwang},
  \citenamefont {Jang}, \citenamefont {Song}, \citenamefont {Brongersma},
  \citenamefont {Ha}, \citenamefont {Hwang}, \citenamefont {Choo},\ and\
  \citenamefont {Choi}}]{park_all-solid-state_2021}%
  \BibitemOpen
  \bibfield  {author} {\bibinfo {author} {\bibfnamefont {J.}~\bibnamefont
  {Park}}, \bibinfo {author} {\bibfnamefont {B.~G.}\ \bibnamefont {Jeong}},
  \bibinfo {author} {\bibfnamefont {S.~I.}\ \bibnamefont {Kim}}, \bibinfo
  {author} {\bibfnamefont {D.}~\bibnamefont {Lee}}, \bibinfo {author}
  {\bibfnamefont {J.}~\bibnamefont {Kim}}, \bibinfo {author} {\bibfnamefont
  {C.}~\bibnamefont {Shin}}, \bibinfo {author} {\bibfnamefont {C.~B.}\
  \bibnamefont {Lee}}, \bibinfo {author} {\bibfnamefont {T.}~\bibnamefont
  {Otsuka}}, \bibinfo {author} {\bibfnamefont {J.}~\bibnamefont {Kyoung}},
  \bibinfo {author} {\bibfnamefont {S.}~\bibnamefont {Kim}}, \bibinfo {author}
  {\bibfnamefont {K.-Y.}\ \bibnamefont {Yang}}, \bibinfo {author}
  {\bibfnamefont {Y.-Y.}\ \bibnamefont {Park}}, \bibinfo {author}
  {\bibfnamefont {J.}~\bibnamefont {Lee}}, \bibinfo {author} {\bibfnamefont
  {I.}~\bibnamefont {Hwang}}, \bibinfo {author} {\bibfnamefont
  {J.}~\bibnamefont {Jang}}, \bibinfo {author} {\bibfnamefont {S.~H.}\
  \bibnamefont {Song}}, \bibinfo {author} {\bibfnamefont {M.~L.}\ \bibnamefont
  {Brongersma}}, \bibinfo {author} {\bibfnamefont {K.}~\bibnamefont {Ha}},
  \bibinfo {author} {\bibfnamefont {S.-W.}\ \bibnamefont {Hwang}}, \bibinfo
  {author} {\bibfnamefont {H.}~\bibnamefont {Choo}},\ and\ \bibinfo {author}
  {\bibfnamefont {B.~L.}\ \bibnamefont {Choi}},\ }\href
  {https://doi.org/10.1038/s41565-020-00787-y} {\bibfield  {journal} {\bibinfo
  {journal} {Nature Nanotechnology}\ }\textbf {\bibinfo {volume} {16}},\
  \bibinfo {pages} {69} (\bibinfo {year} {2021})},\ \bibinfo {note} {publisher:
  Nature Publishing Group}\BibitemShut {NoStop}%
\end{thebibliography}
